\documentclass[twocolumn]{aa}
\usepackage{psfig}
\usepackage{natbib}
\bibpunct{(}{)}{;}{a}{}{,}
\usepackage{ltxtable}
\usepackage[latin1]{inputenc}
\begin{document}
   \title{Temporal evolution of magnetic molecular shocks}
  \subtitle{I. Moving grid simulations}

   \author{P. Lesaffre \inst{1,2,3}         
        \and
        J.-P. Chièze \inst{3}	
	\and
	S. Cabrit \inst{4}
	\and
	G. Pineau des Forêts \inst{5,6}
	  }

   \offprints{lesaffre@ast.cam.ac.uk}

   \institute{Institute of Astronomy, Madingley Road, Cambridge 
CB3~0HA, UK 
\and University of Oxford, Department of Astrophysics,
 Oxford OX1~3RH, UK
\and CEA/DAPNIA/SAp Orme des Merisiers, F-91191
   Gif-sur-Yvette Cedex 
\and LERMA, UMR 8112 du CNRS, Observatoire de
   Paris, 61 Av. de l'Observatoire, F-75014 Paris 
\and IAS, UMR-8617 du CNRS, Université Paris-Sud, bât. 121, F-91405 Orsay 
\and LUTH,
   UMR-8102 du CNRS, Observatoire de Paris, F-92190 Meudon Cedex }
   \date{Received September 15, 1996; Accepted March 16, 1997}

   \abstract{ We  present time-dependent 1D simulations of multifluid magnetic
   shocks with chemistry  resolved down to the mean free path. They are 
   obtained with an adaptive moving grid  implemented with an
   implicit scheme. We examine a broad range of parameters relevant
   to conditions in dense molecular clouds, with preshock densities
   $10^3$ cm$^{-3}<n<10^5$~cm$^{-3}$, velocities $10$ kms$^{-1}$$~<u<40$
   kms$^{-1}$, and three different scalings for the transverse
   magnetic field~: $B=0,0.1,1~\mu$G$\times \sqrt{n/{\rm cm}^{-3}}$.

   We first use this study to validate the results of Chièze, Pineau
   des Forêts \& Flower (1998), in particular the long delays
   necessary to obtain steady C-type shocks, and we provide
   evolutionary time-scales for a much greater range of parameters.

  We also present the first time-dependent models of dissociative shocks
  with a magnetic precursor, including the first models of stationary CJ
  shocks in molecular conditions.  We find that the maximum speed for
  steady C-type shocks is reached {\it before} the occurrence of a sonic
  point in the neutral fluid, unlike previously thought.  As a result,
  the maximum speed for C-shocks is lower than previously believed.

  Finally, we find a large amplitude bouncing instability in J-type fronts
  near the H$_2$ dissociation limit ($u\simeq 25-30~$kms$^{-1}$), driven by H$_2$
  dissociation/reformation.  At higher speeds, we find an
  oscillatory behaviour of short period and small amplitude linked to
  collisional ionisation of H. Both instabilities are
  suppressed after some time when a magnetic field is present.
  
  In a companion paper, we use the
  present simulations to validate a new semi-analytical construction
  method for young low-velocity magnetic shocks based on truncated
  steady-state models.

\keywords{magnetohydrodynamics -- interstellar matter -- shock wave
   propagation -- time dependence -- hydrogen -- molecular oscillations}}

   \maketitle

\section{Introduction}

 Today, essentially all observational diagnostics of shocks are based
on steady state models. Indeed, the steady state form of the
monodimensional equations of hydrodynamics for these shocks is an ordinary
differential equation. Therefore, computation of such models can encompass
the luxury of details involved in the thermal, magnetic and chemical
richness of the interstellar medium.
\citet{HM79,HM80,HM89} describe the destruction and reformation of
molecules in very strong jump (hereafter J-type) shocks, after having
studied their radiative precursor \citep{SM79}. \citet{M71} discovered
that high magnetic fields can transfer kinetic energy to thermal
energy in a continuous manner (C-type shocks). \citet{D80},
\citet{D83}, and \citet{RD90} describe the multifluid nature of these
shocks.  In a series of papers \citet{F85,F86,FP95,L02,F03, FP03}
carefully examine the relevant chemistry, and in a more general way,
all collisional interactions between the charged and neutral
fluids. They point out the strong influence of chemistry on the
magnetic precursor, carefully address the energy losses linked with
this chemistry, and stress the need to follow the level populations of
H$_2$ along the flow. H$_2$ is indeed one of the most efficient
coolants, and therefore one of the most powerful diagnostics. A good
review of the steady state J-type and C-type models can be found in
\citet{DM93}.  All these studies assume a purely transverse
magnetic field frozen in the ions. Other authors
\citep{P90, PH94, C97, W98} have investigated the dynamical role of the
grains as well as the effect of oblique magnetic fields. For high pre-shock
densities, they showed that the drift velocity had been underestimated,
producing a significant rise in the maximum temperature.  They also
discovered that rotation of the magnetic field could bring one of the
charged fluids at rest in the shock frame.

Time-dependent studies of dense shocks in one dimension have focused
separately on  three points. The first one is the oscillatory
instabilities due to the temperature dependence of the cooling of the gas
\citep{CI82,G88,SB95,WF96,Lim02,SR03}. The second one is the early
evolution of multifluid shocks, which is found to involve a combination of
C and J shock fronts \citep{P97,SM97,CPF98}. \citet{P97} and \citet{CPF98}
include non-equilibrium cooling and chemistry in the fluids, with a degree
of details comparable to steady state models. They prove that in low
ionised magnetised media, the steady state could be reached only after very
late ages (up to a few times 10$^5$ yr), far greater than the variation time
scales of shock driving sources. The third point is the dynamical
effects of grains, which is beginning to be included in time-dependent
studies \citep{CR02,Fal03}.

   Chièze, Pineau des Forêts \& Flower (1998) used a time
  stretching method to resolve the sharp J-fronts in time-dependent
  magnetised shocks. However, their method did not allow to resolve
  non-viscous fronts (e.g. H$_2$ reformation regions), and could
  introduce synchronisation problems. To overcome both limitations, we
  have developed a new moving grid algorithm which allows one to resolve
  discontinuities down to the mean free path.  Although we do not
  include the treatment of grain dynamics to remain consistent
  with \citet{CPF98}, we note that our numerical scheme already provides
  the required framework to solve the problems mentioned by
  \citet{Fal03}, as it is implicit {\it with} very high resolution. As
  such, it is the first algorithm able to model time-dependent
  multifluid dissociative shocks.

  In this paper, we use our code to validate the evolutionary
  behaviour and time scales obtained with the ''anamorphosis'' method
  of \citet{CPF98}, and to extend their range of shock parameters to
  cover the denser conditions encountered in protostellar jets and
  molecular clouds, including dissociative and partly ionising
  shocks. We (1) compute the evolution time-scales of these shocks,
  and interpret physically their behaviour, (2) find that stationary
  CJ shocks are obtained before the occurence of a sonic point in the
  shock frame and are thus more frequent than previously thought, (3)
  describe two types of oscillations driven by chemistry that were not
  identified previously.
 
 In Sect. 2, we briefly present our numerical model, with the
improvements made compared to \citet{CPF98}.  Section 3 describes the
results and compares them to previous work. Section 4 discusses the
limitations of our method and Sect. 5 summarises our conclusions.


\section{Numerical and physical inputs}

\subsection{Numerical scheme}

\subsubsection{Time-step integration}
 
  We use a fully non-linear implicit scheme for time integration. The
  implicitation parameter is set equal to 0.55~: that way, we combine a
  quasi order 2 accuracy in time with an unconditional stability of the
  scheme. We use a Van-Leer advection scheme. The relative variations
of the variables is kept under 5\% at each time-step.
We reproduce classical tests
  such as the Sod's shock tube, Rankine-Hugoniot relations, and Sedov
  explosion with a 0.3\% accuracy.
 
 More details can be found in \citet{Les02}.

  We solve for the same equations as \citet{CPF98}, although the
  hydrodynamics and chemistry are now solved simultaneously.    This
  amounts to a number of 33 chemical equations, 2 momentum equations,
  4 energy equations and one equation for the moving mesh all coupled
  together.  Solving for the chemistry without splitting it from the
  hydrodynamics helps numerical convergence, but has a computational
  cost, since the  Jacobian matrix is far heavier to be
  inverted. In a few cases, the code stalled because a proper
  convergence of the Newton-Raphson iterations required a time step
  much too small.

\subsubsection{Adaptive grid}
  Chemistry encompasses an extensive range of time scales, down to
  times as short as a few mean collision times of the particles. When
  this range of time scales is coupled with hydrodynamics, it
  generates a full range of spatial scales, down to a few mean free
  paths.

  To achieve this extremely high resolution where needed, we use the
  moving grid algorithm designed by \citet{DD87}. We define the delay
  time parameter of the grid as the sound crossing time of the
  smallest zone. Our resolution
  function combines three arguments~: the gradient of the neutral
  temperature, the gradient of the ionic temperature, and the maximum
  of all chemical gradients. 

  Thanks to this method, we were able to resolve all shocks and
  chemical fronts with  a total number of zones as low as 100,
  without having to resort to the anamorphosis method of
  \citet{CPF98}.  The present code can safely handle two or more
  sharp features like a strong shock followed by a molecular
  reformation region, or a shock, a contact discontinuity and a
  reverse shock.  The natural viscosity is used in the shock
  fronts, resulting in a viscous spread of typically
  $10^{-3}$~pc/($n_{\rm H}$.cm$^3$) resolved with about 10 zones.

\subsubsection{Diffusion}

We consider diffusion for the chemical species, based on a Fick law with a
diffusion length equal to the local mean free path $
\lambda_{\rm d} =1/\sigma n_{\rm H}$ where
$\sigma=10^{-15}~$cm$^2$ is the molecular cross section, and $n_{\rm H}$ is
the numerical density of hydrogen nuclei. This is motivated by
the fact that in regions where molecules are reforming, the fluid
velocities are usually very low. The reformation times of the
molecules would then lead to an unrealistically  thin front. This
diffusion term ensures that the front has at least a width of the
order of the mean free path.

\subsection{Physical inputs}

\subsubsection{Chemistry}

The network comprises around 120 reactions involving 33 different
species (including electrons).   It is the same network as
\citet{CPF98} with a few additional (or updated) reactions~:
\begin{itemize}
 \item collisional dissociations of H$_2$ by e$^-$ \citep{F96}, H, H$_2$, He,
 and streaming ions \citep{Wal00}.

 \item collisional ionisation of H by e$^-$ \citep{L02}.

 \item updated H$_2$ formation on grains~: for the sticking coefficient, we
 use expression (4) of \citet{HS71} calibrated by \citet{BZ91} and
\citet{M98}.
\end{itemize}

  To compute the energy transfers and coolings due to chemical reactions,
we properly track the fluid to which each species belong. This is critical
especially for endothermic reactions like collisional ionisation of H.

\subsubsection{Atomic and molecular cooling}
\label{cooling}
 
  We use refined versions of all the atomic and molecular coolings
  used by \citet{CPF98}, with the addition of H cooling.

 At each time-step, we compute the level population of C (4 levels), C$^+$
 (4 levels) and O (5 levels) excited by e$^-$, H, He, and H$_2$. 
Collisional excitation coefficients are taken from
\citet{M83,HN84,MF87,HM89,L00}.
Lyman $\alpha$ cooling is taken from \citet{F96}, and cooling due to
radiative recombination of H is from \citet{S78} (case B).   We properly
track the energy lost by each fluid, depending on the collider.

 We consider cooling by H$_2$ \citep{L99}, OH
\citep{HM79}, CO and H$_2$O \citep{NK93}. Since the latter authors
provide data for excitation of CO by H$_2$ only, we assumed the same
excitation coefficients for H-CO collisions. The optical depth
parameter is computed in the static approximation, with a distance
scale corresponding to $A_{\rm v}$ = 10 mag.

\subsection{Parameter space and initial conditions } 

The three parameters $n$ (number density of hydrogen nuclei), $u$ (initial
velocity), and $B$ (initial ambient magnetic field) characterise a
simulation. We replace $B$ by the parameter $b$, setting
$B=b(n.$cm$^3)^\frac12~\mu$G. Typical conditions in the interstellar medium
correspond to $b=1$. But as we consider only the transverse component of
the field, $b$ can take values between 0 and 1. We combine $n=10^3$,
$10^4$, $10^5$~cm$^{-3}$, $u=10$, 20, 30, 40~kms$^{-1}$, and $b=0$, 0.1, 1 and
produce a grid of 36 different simulations. To test the dynamical
instability that we observed in dissociative cases, we ran two additional
cases with $b=0$, $u=25$~kms$^{-1}$, $n=10^4$ and $10^5$~cm$^{-3}$. The abundances
of He, C, O, and Fe nuclei relative to $n_{\rm H}$ are the following~:
He=0.1 ; C=1.7$~10^{-4}$ ; O=4.25~$10^{-4}$ ; Fe=1.8$~10^{-7}$.\\

We use a piston-like protocol. The initial conditions of the simulation box
are homogeneous, in thermal and chemical equilibrium with a density
corresponding to a given $n_{\rm H}=n$.  The transverse magnetic field is
uniform, equal to $B$. All interfaces of the cells in the simulation have a
velocity $u$, except the last one, which is a fixed reflexive boundary. The
moving grid algorithm allows us to begin with a very small simulation box
of a few mean free paths (typically 30). We make the left border of the box
continuously flee the shock front. That way, the adiabatic shock is always
fully resolved. The simulation is stopped when the box is too large and the
resolution of sharp fronts cannot be supported any more by the
algorithm. This occurs when the dynamical range of length scales is greater
than typically a billion.


\section{Results}

\subsection{Final steady-states and trajectories in the piston frame}

Table \ref{timescales} summarises the final steady-state of each model
as a function of the initial parameters $b$, $n$, and $u$. When $b=0$,
all flows evolve to stable J-type shocks, except at intermediate
velocities for which we find strong or moderate undamped
oscillations. When $b$ is finite, the flow generally evolves to a
steady C-type shock. At sufficiently low $b$ or high $u$, however, a
strong J shock in the neutrals persists behind the continuous magnetic
precursor, and a steady CJ-type shock is obtained.  Figures
1-2-3 show the thermal and
chemical structure of typical shocks in their final J, CJ and C-type
steady-state. Figures 4-5 show intermediate states of J-shocks with higher
entrance velocities.

\begin{figure*}[h]
\begin{tabular}{ccc}
{\bf Fig.~1~: J-shock} & {\bf Fig.~2~: CJ-shock} & {\bf
Fig.~3~: C-shock}\\
\psfig{file=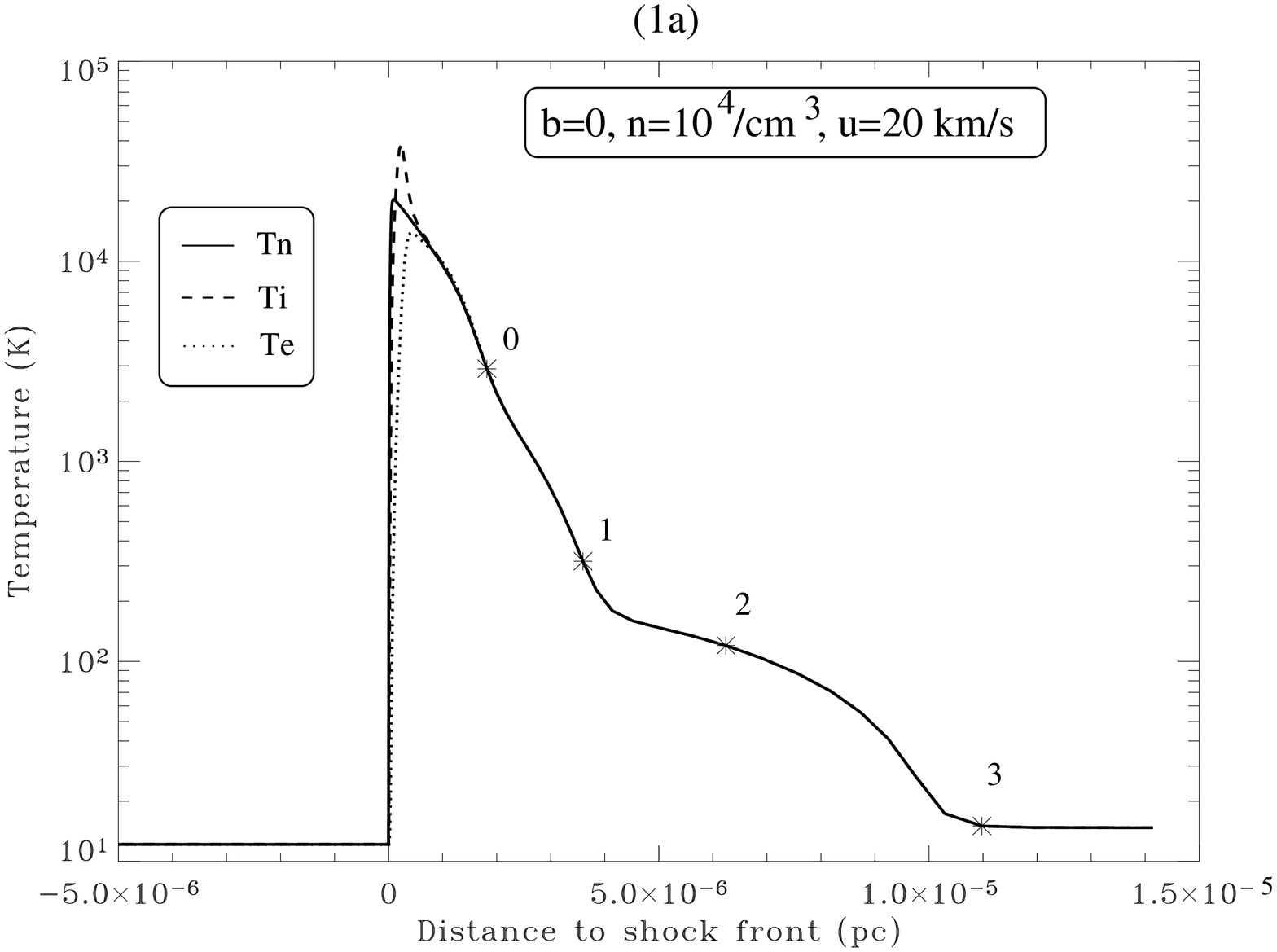,width=6cm,angle=-90}&
\psfig{file=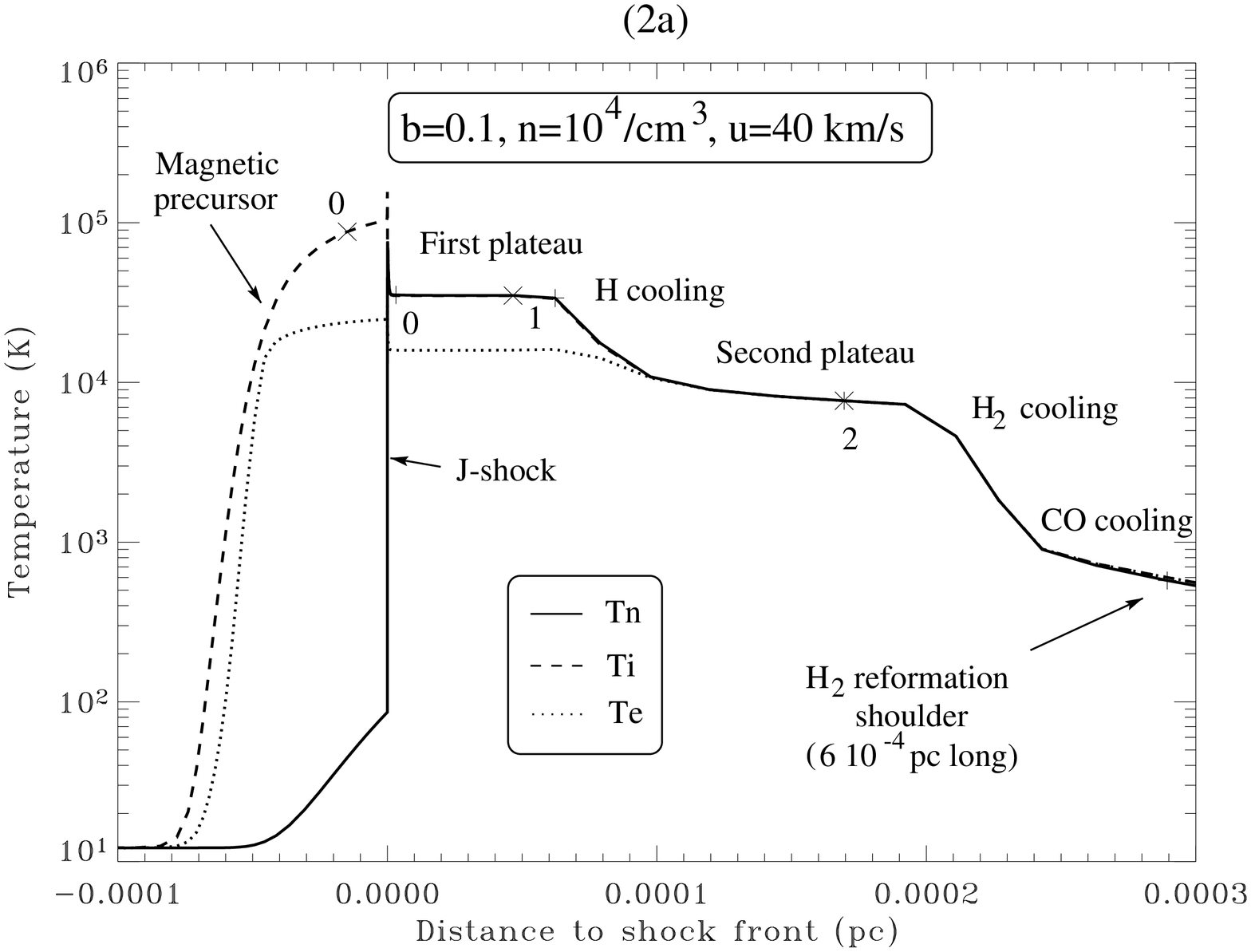,width=6cm,angle=-90}&
\psfig{file=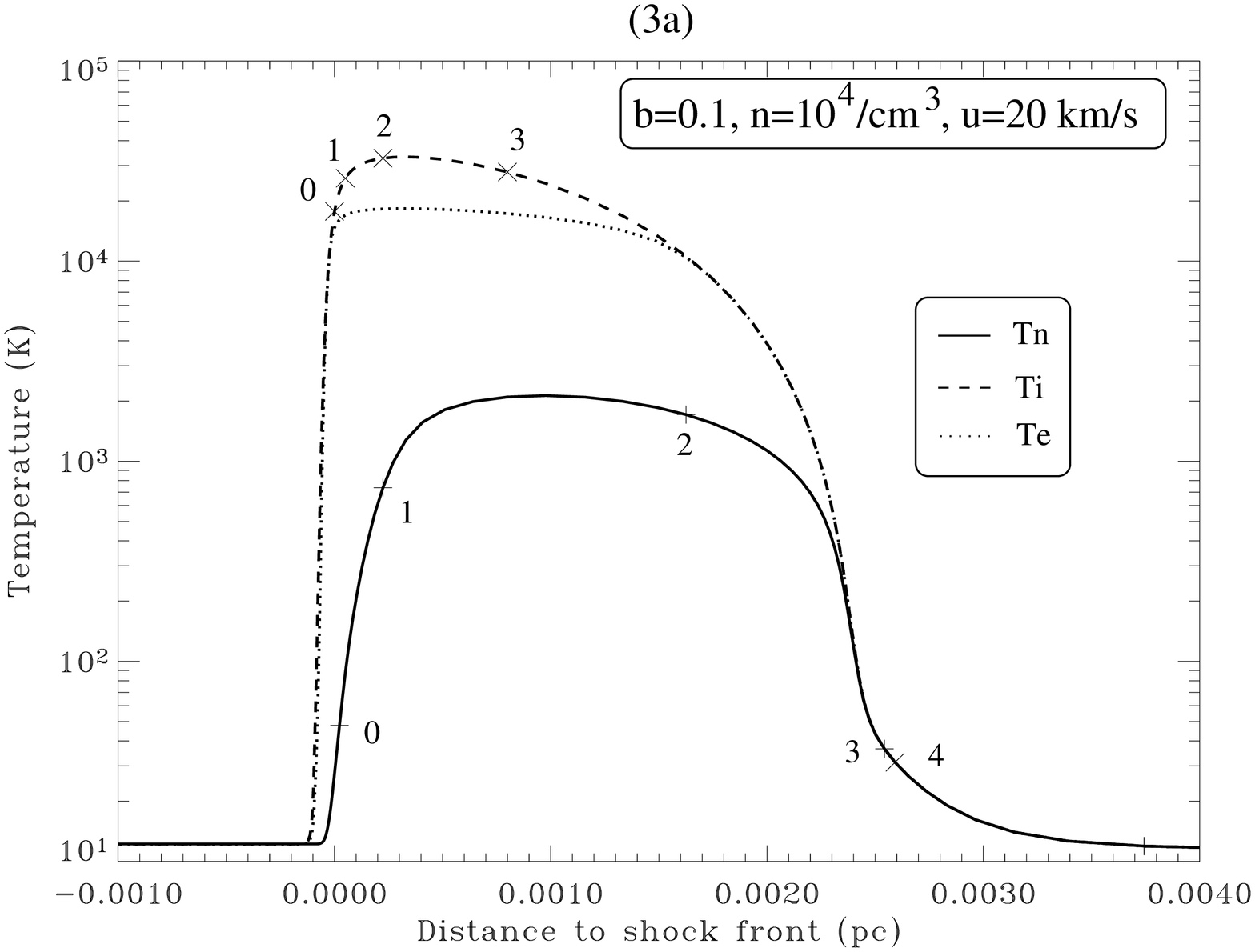,width=6cm,angle=-90}\\
\psfig{file=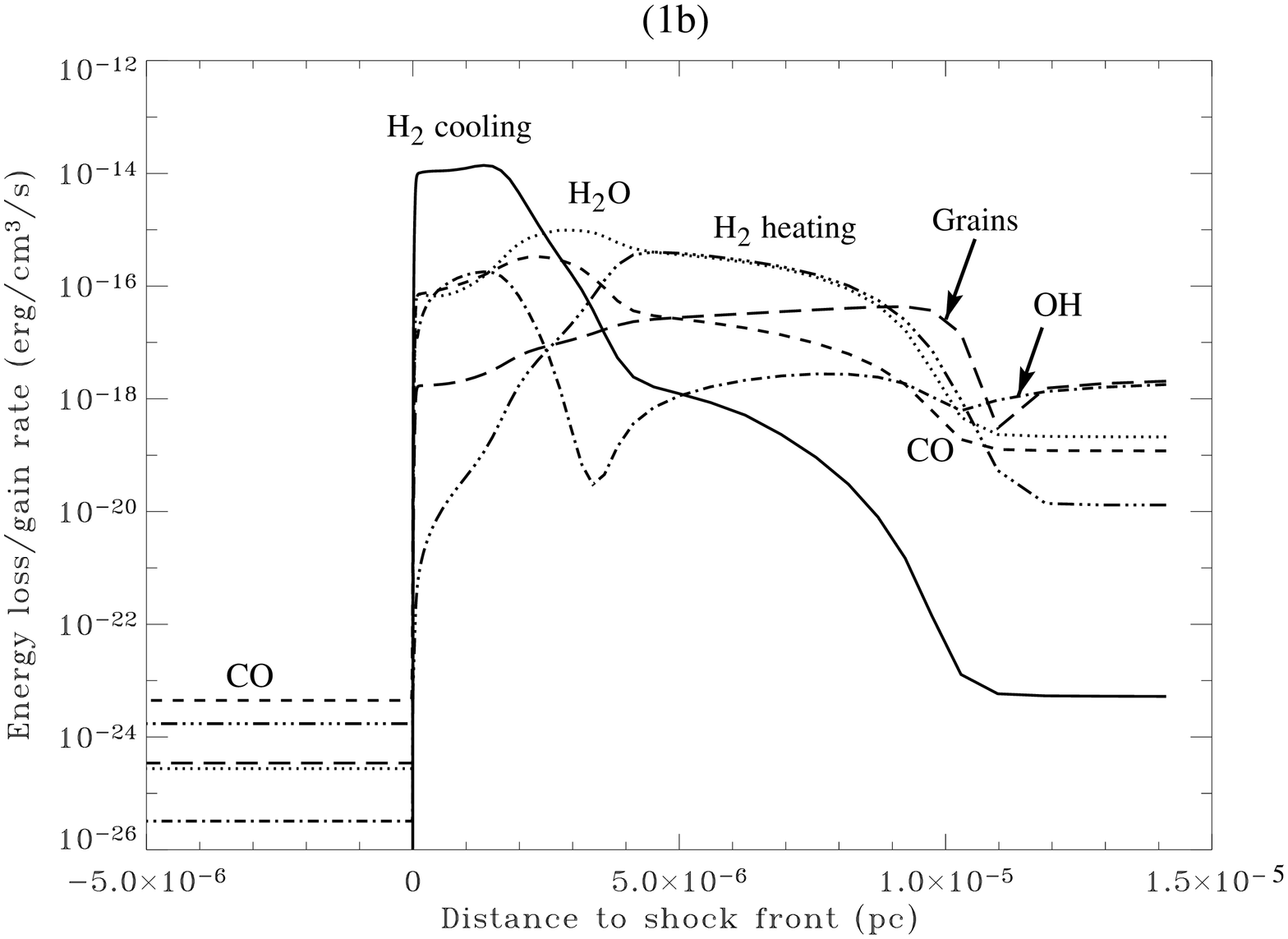,width=6cm,angle=-90}&
\psfig{file=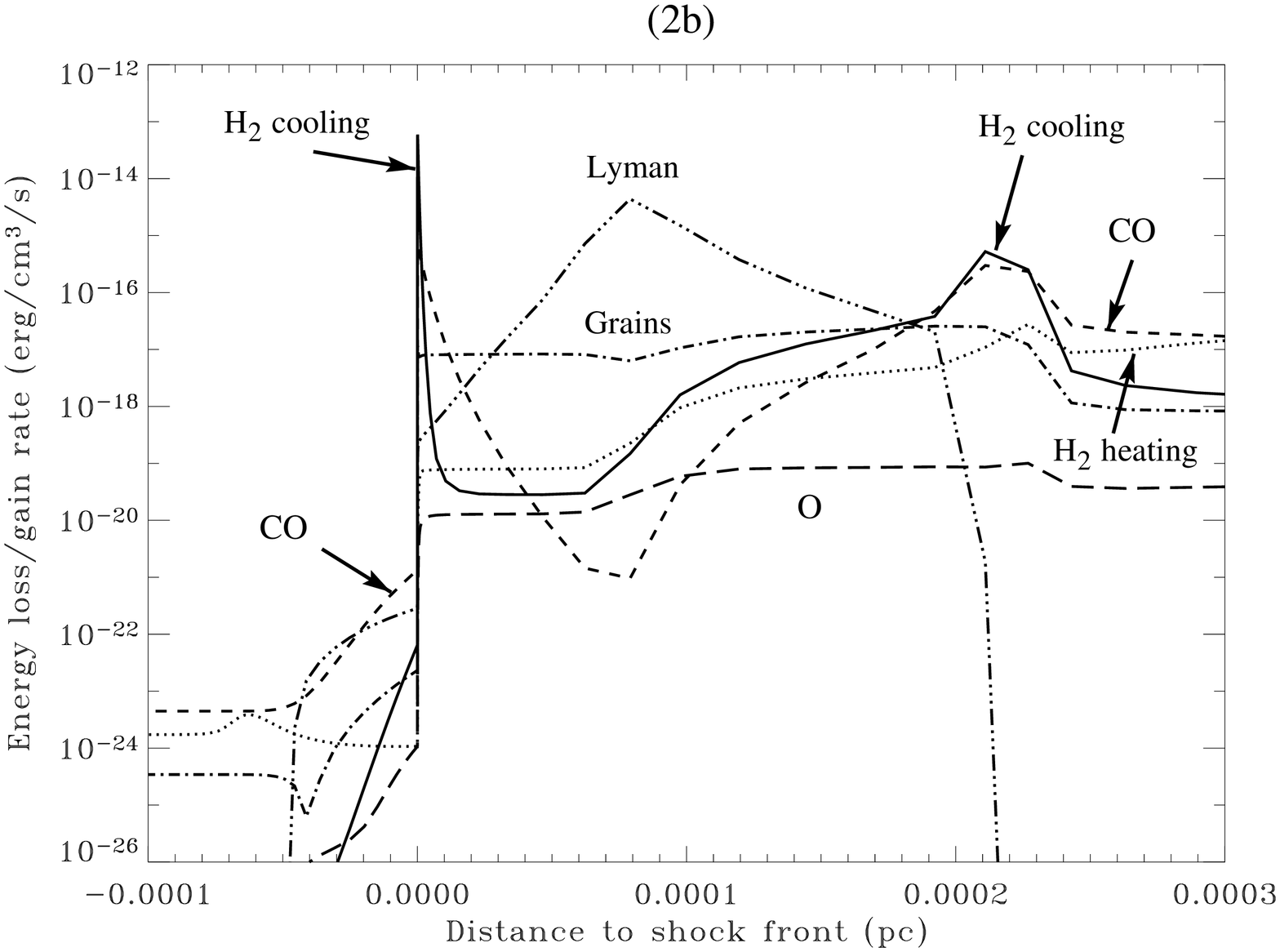,width=6cm,angle=-90}&
\psfig{file=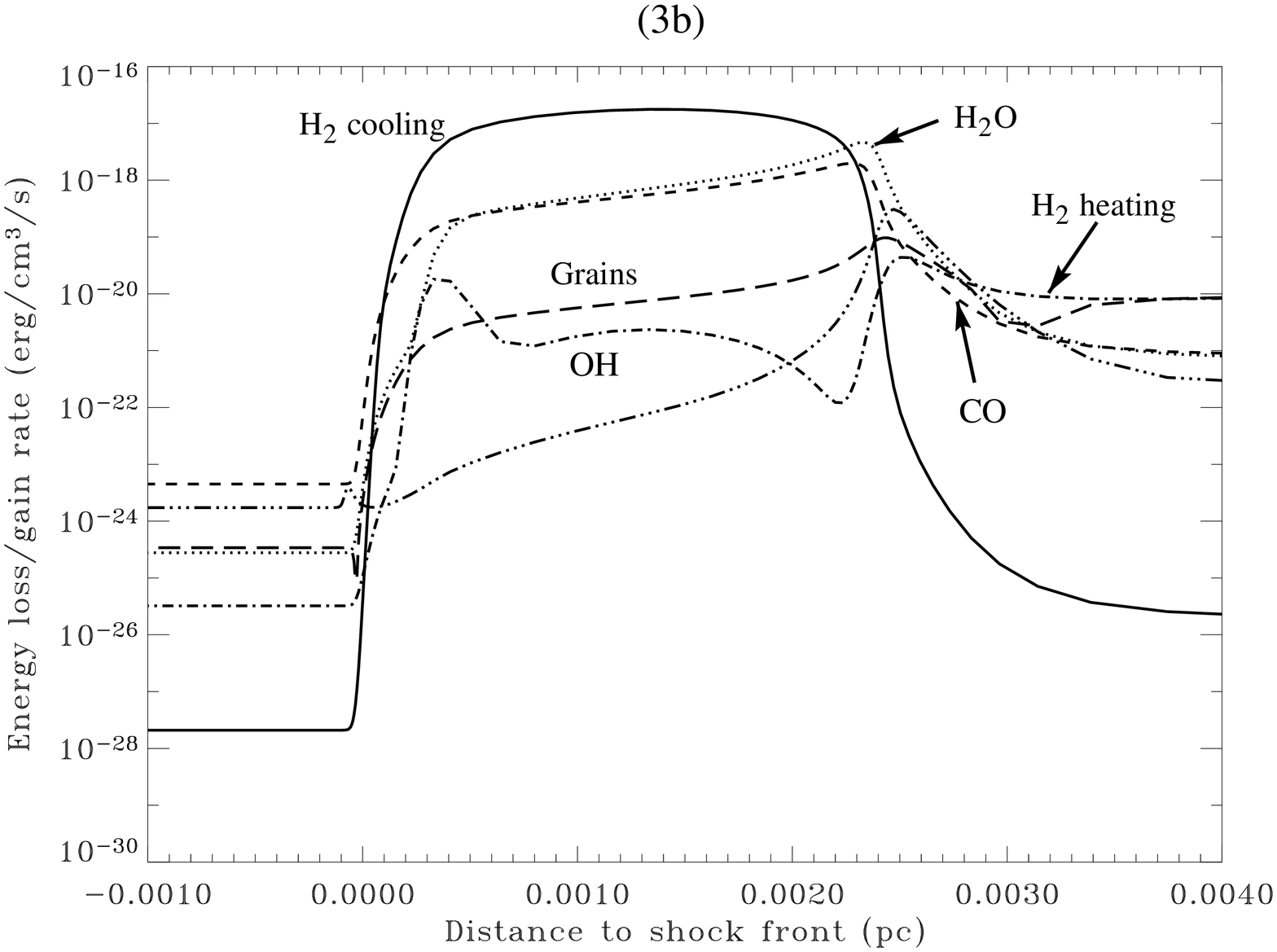,width=6cm,angle=-90}\\
\psfig{file=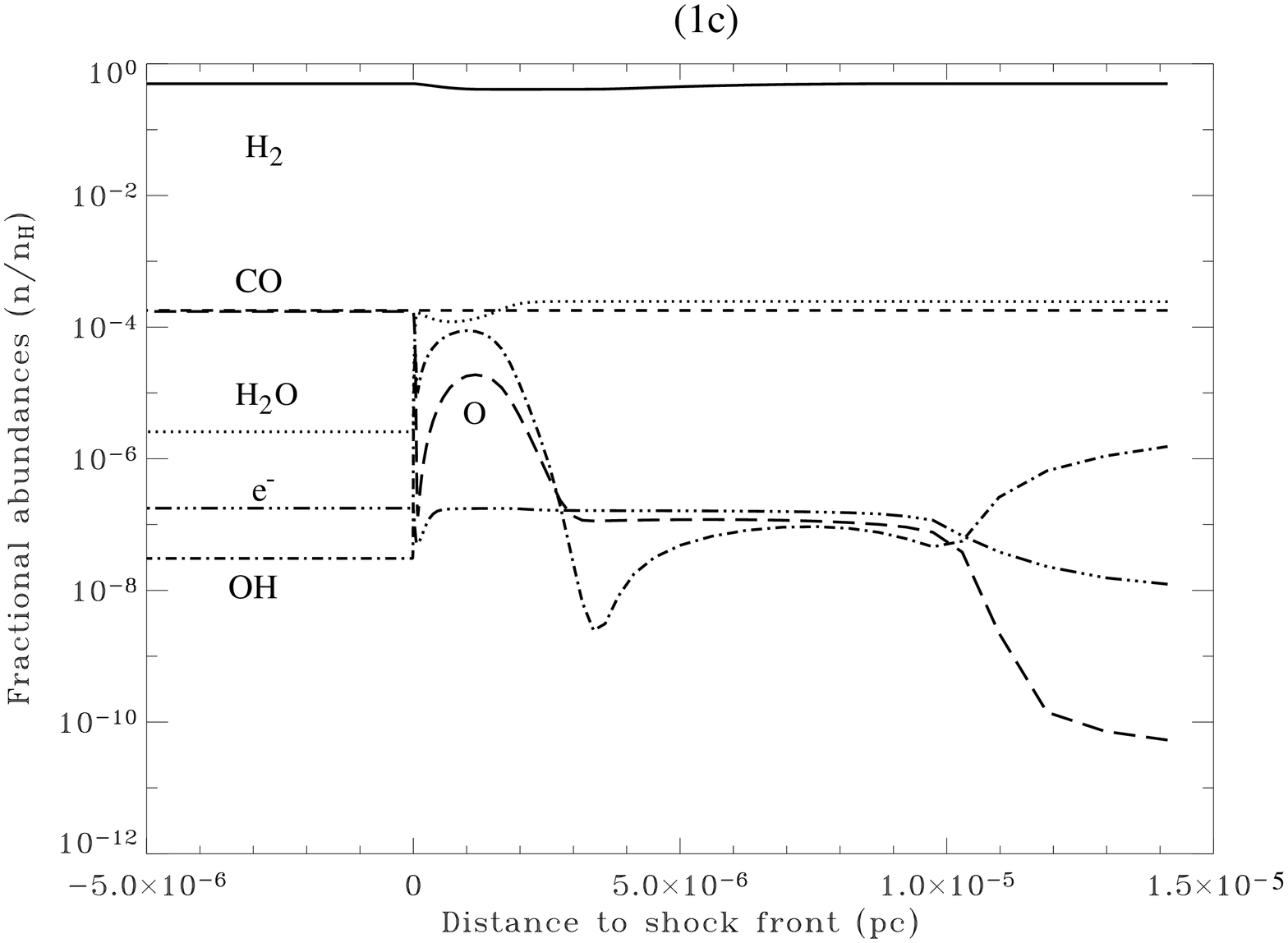,width=6cm,angle=-90}&
\psfig{file=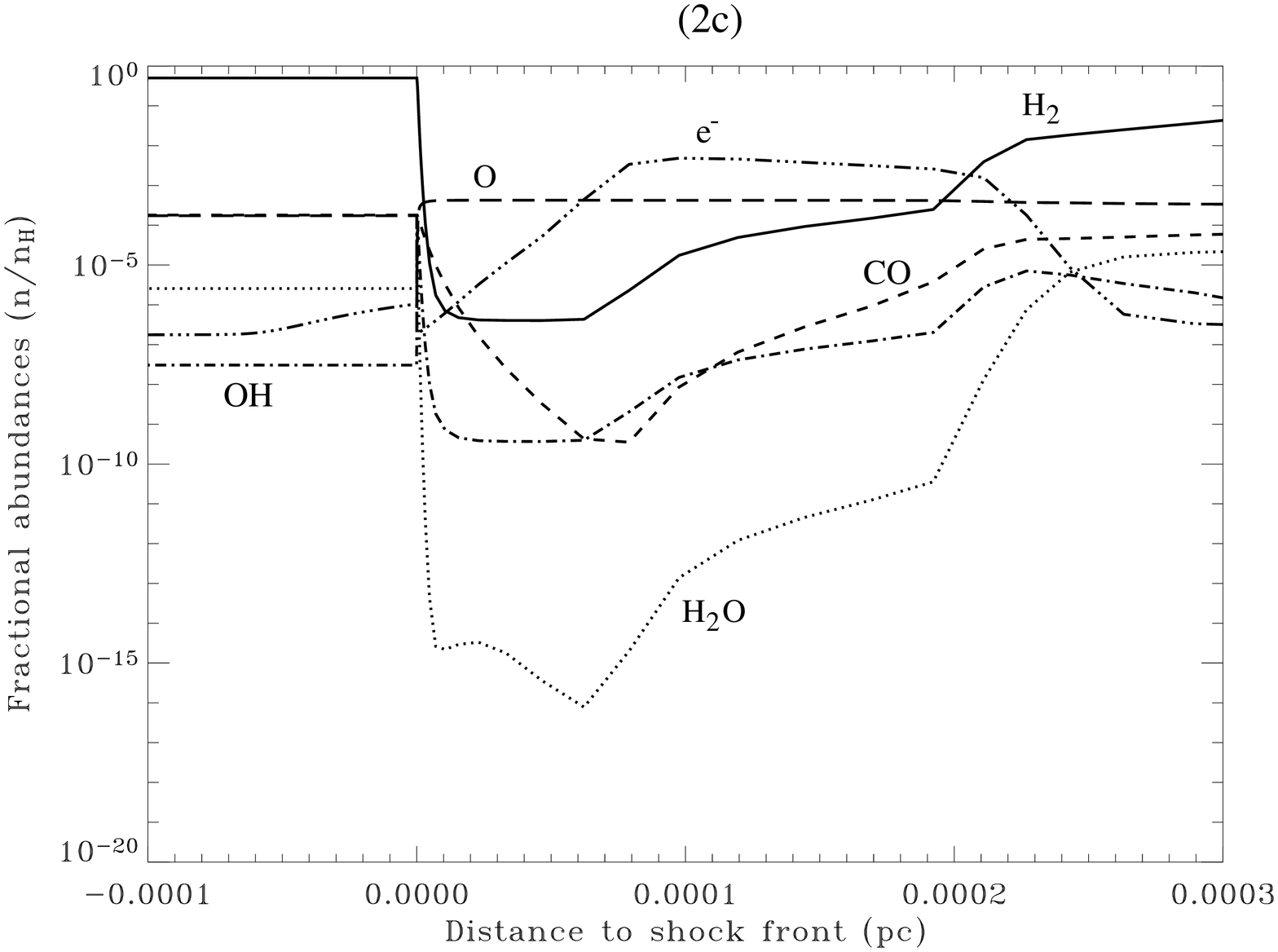,width=6cm,angle=-90}&
\psfig{file=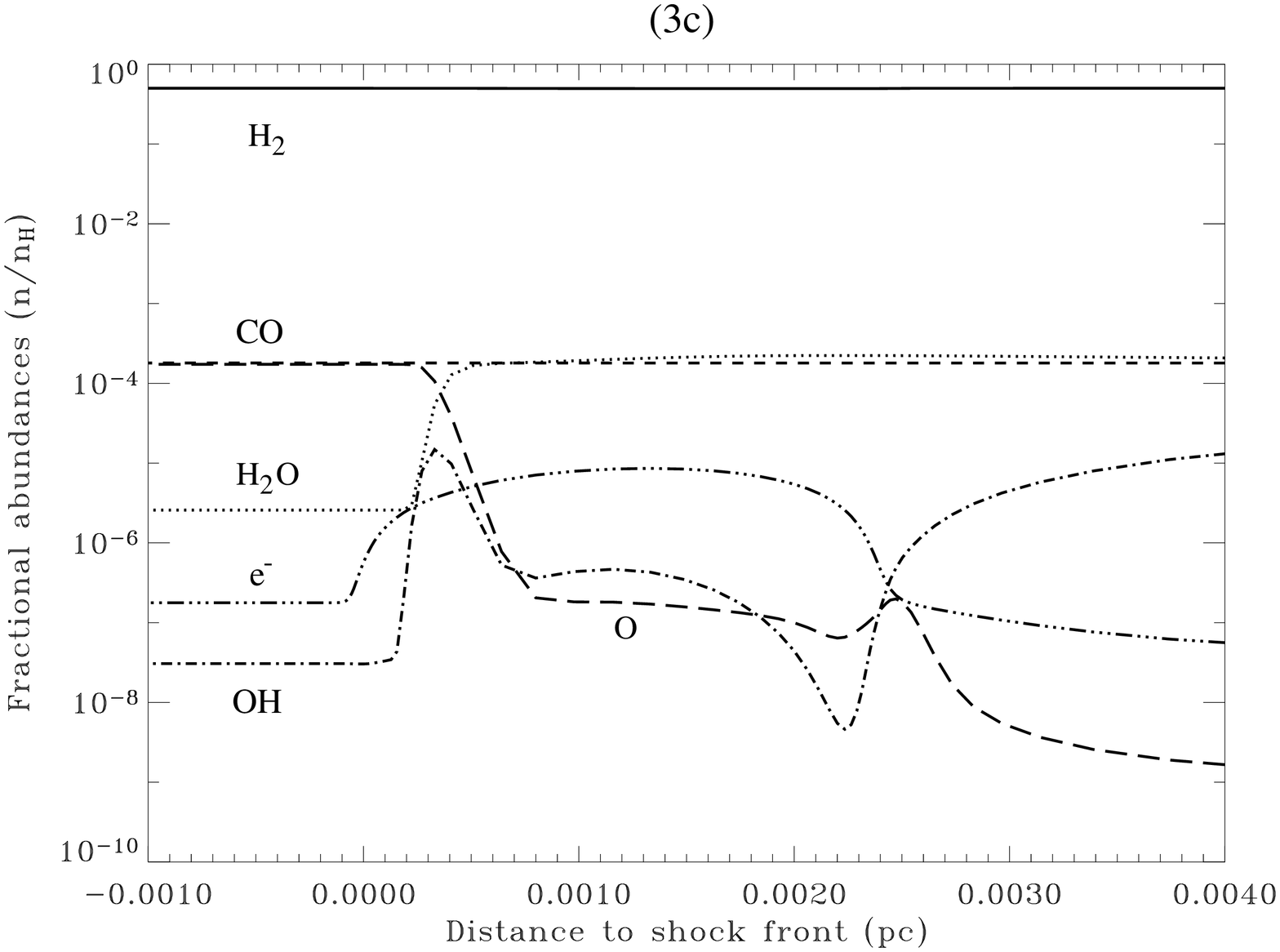,width=6cm,angle=-90}\\
\psfig{file=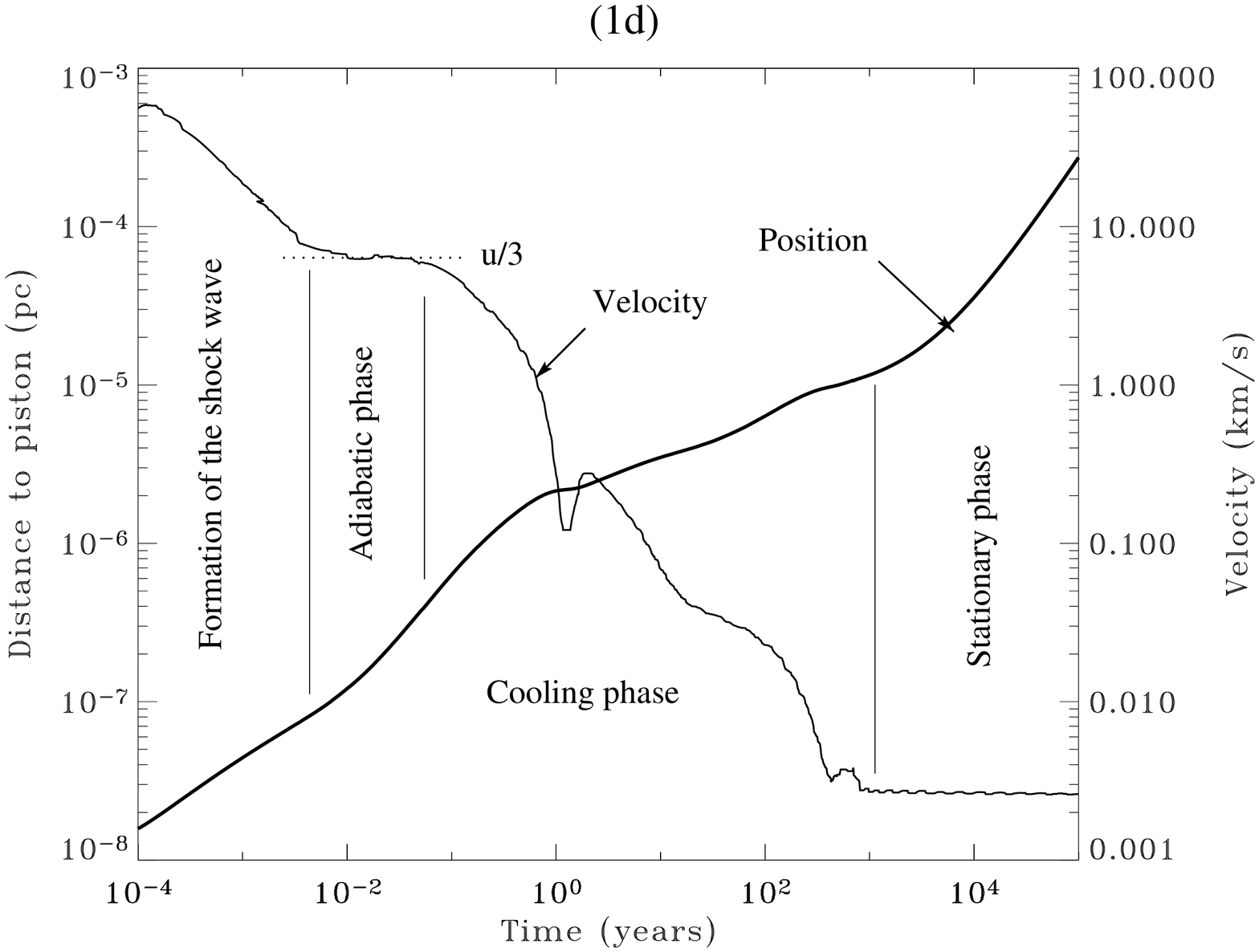,width=6cm,angle=-90}&
\psfig{file=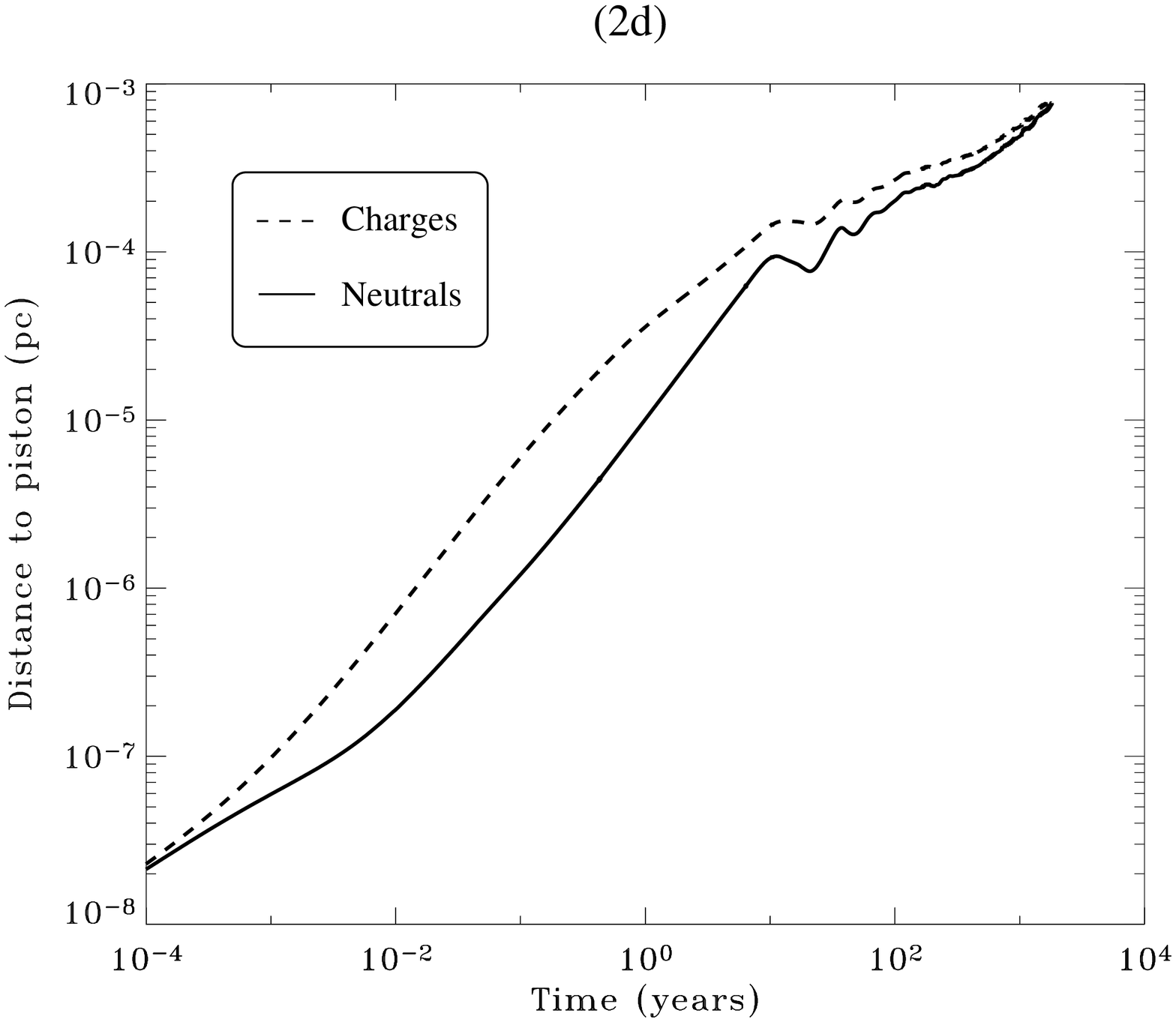,width=6cm,angle=-90}&
\psfig{file=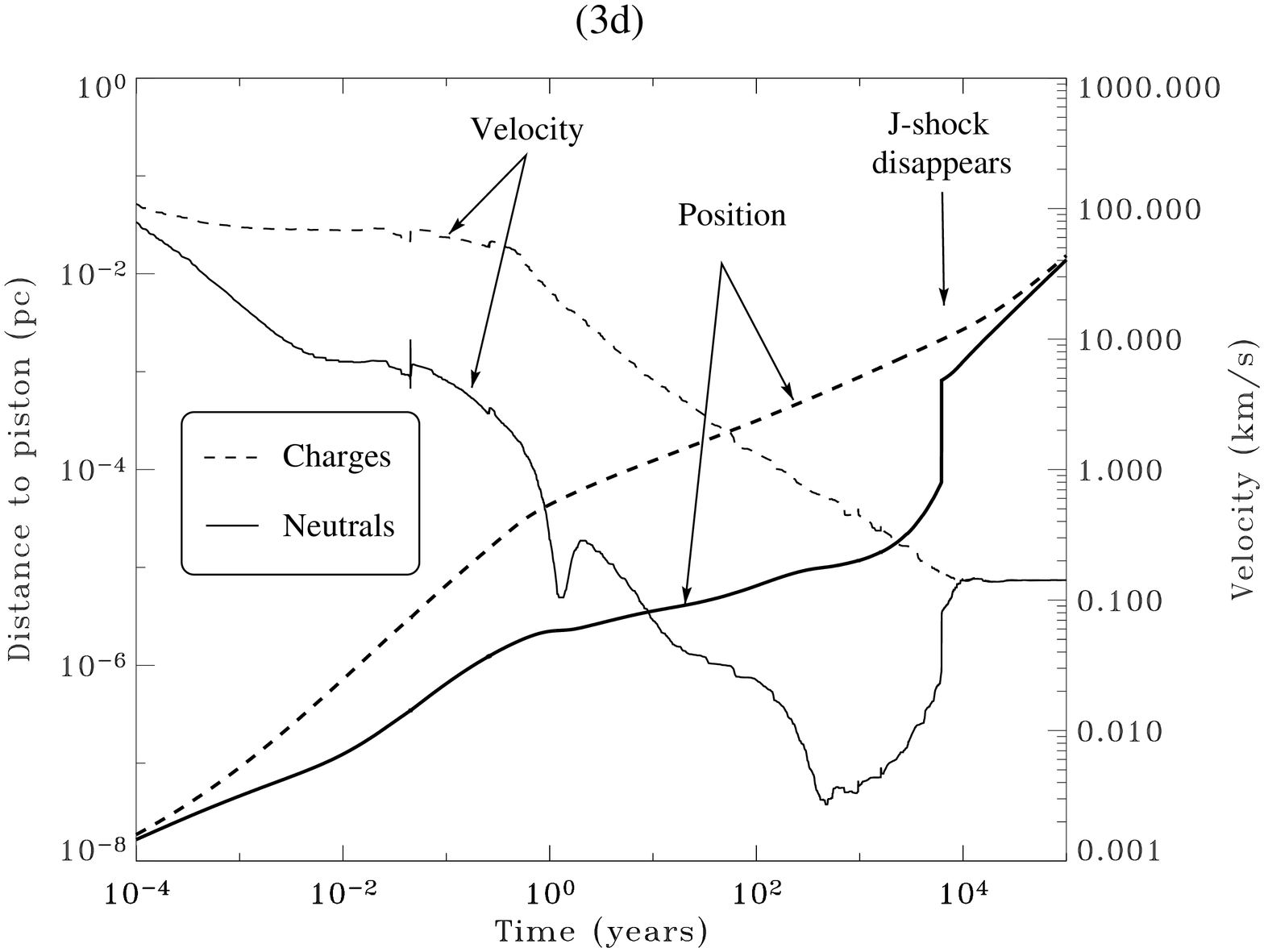,width=6cm,angle=-90}
\end{tabular}
\caption{Stationary structure at $t=2~10^3$~yr and shock trajectories of a J-type shock
with $n=10^4~$cm$^{-3}$, $u=20$~kms$^{-1}$, and $b=0$. (a) {\it Thermal structure~:} temperatures of the
neutrals, ions and electrons. The crosses on the
curve are the points where the flow time computed in the frame of the shock
is equal to $10^n$ yr, $n$ being the number indicated near the
cross. '+' signs stand for neutrals and 'x' for charges.  (b) {\it Cooling
structure~:} magnitude of the main cooling processes.  (c) {\it Chemical
structure~:} abundance relative to hydrogen nuclei of species of interest.
(d) {\it Shock trajectory~:} Position (thick curve) and velocity (thin
curve) of the J-front against time, in the frame of the piston.
{\bf Fig.~2.} Same as Fig.~1 for a steady CJ-type shock with $n=10^4~$cm$^{-3}$,
$u=40$~kms$^{-1}$, and $b=0.1$, at $t=2~10^3$~yr. The dashed curve
in (d) plots the position of the C-precursor against time.
{\bf Fig.~3.} Same as Fig.~2 for a steady C-type shock with $n=10^4~$cm$^{-3}$,
$u=20$~kms$^{-1}$, and $b=0.1$, at $t=10^5$~yr.
}
\label{CJplot}
\end{figure*}
\addtocounter{figure}{2}

\begin{figure*}[h]
\centering{ 
\begin{tabular}{cc}
{\bf Fig.~4~: weakly dissociative J-shock} & {\bf Fig.~5~: partly ionising J-shock} \\
\psfig{file=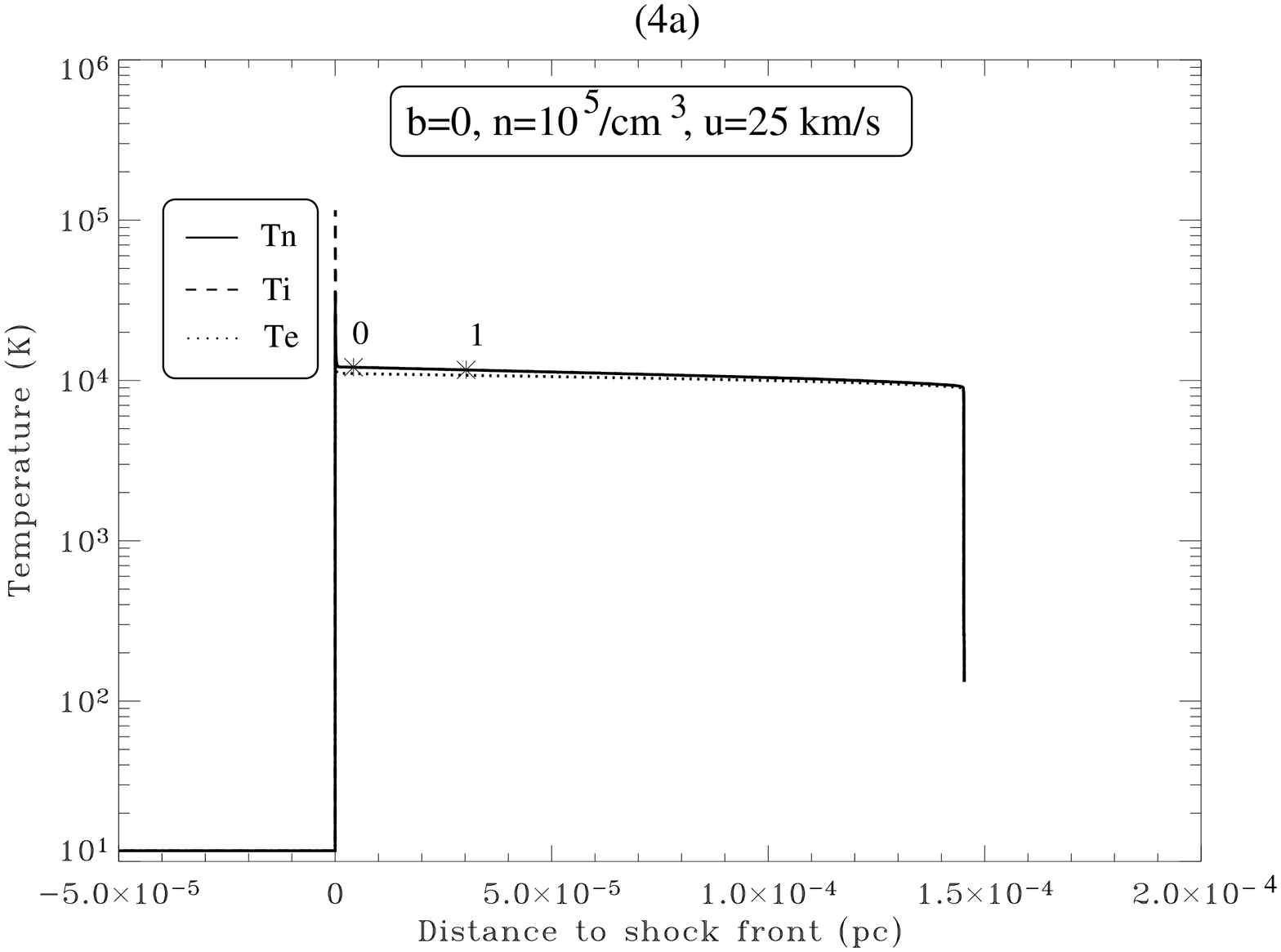,width=6cm,angle=-90}&
\psfig{file=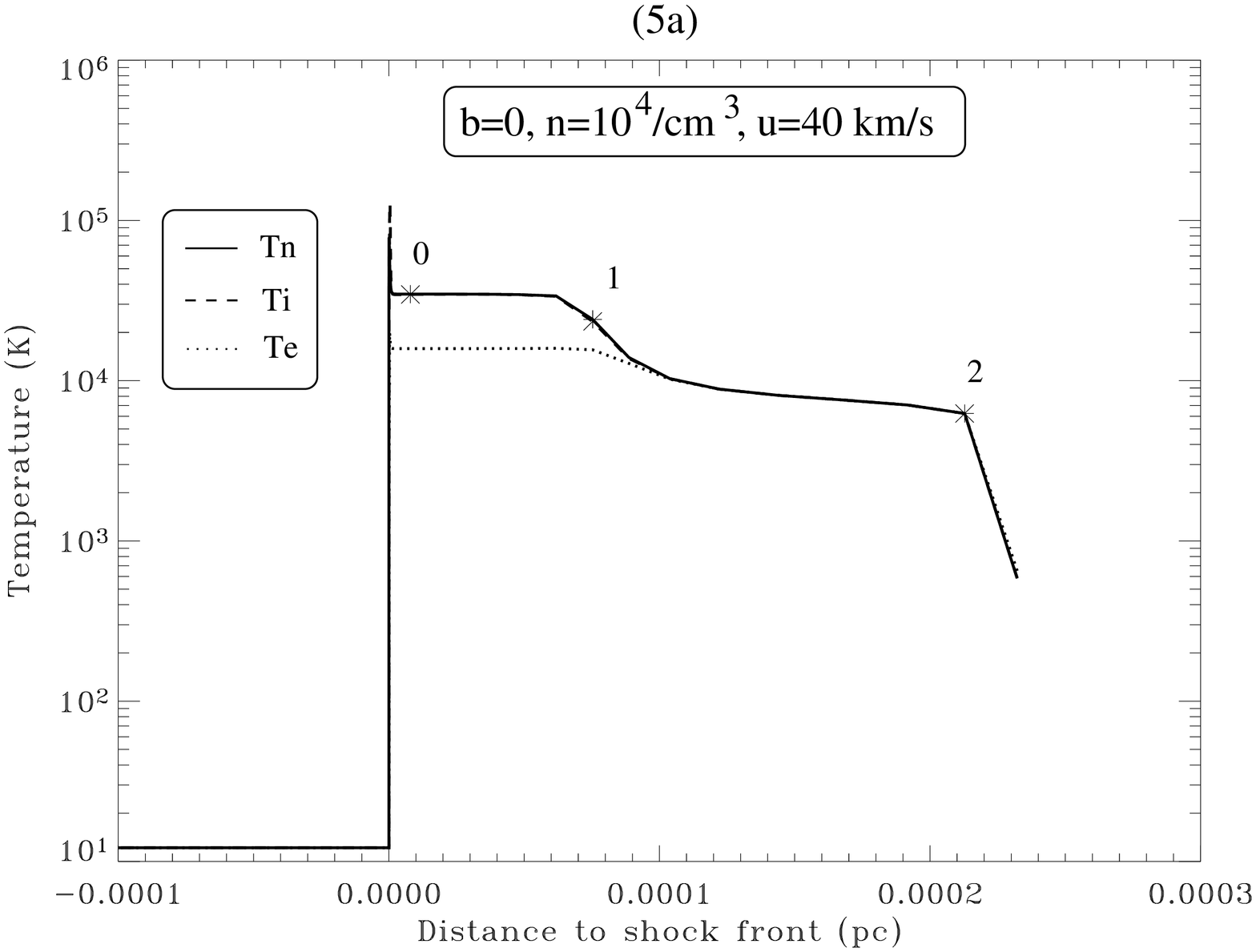,width=6cm,angle=-90}\\
\psfig{file=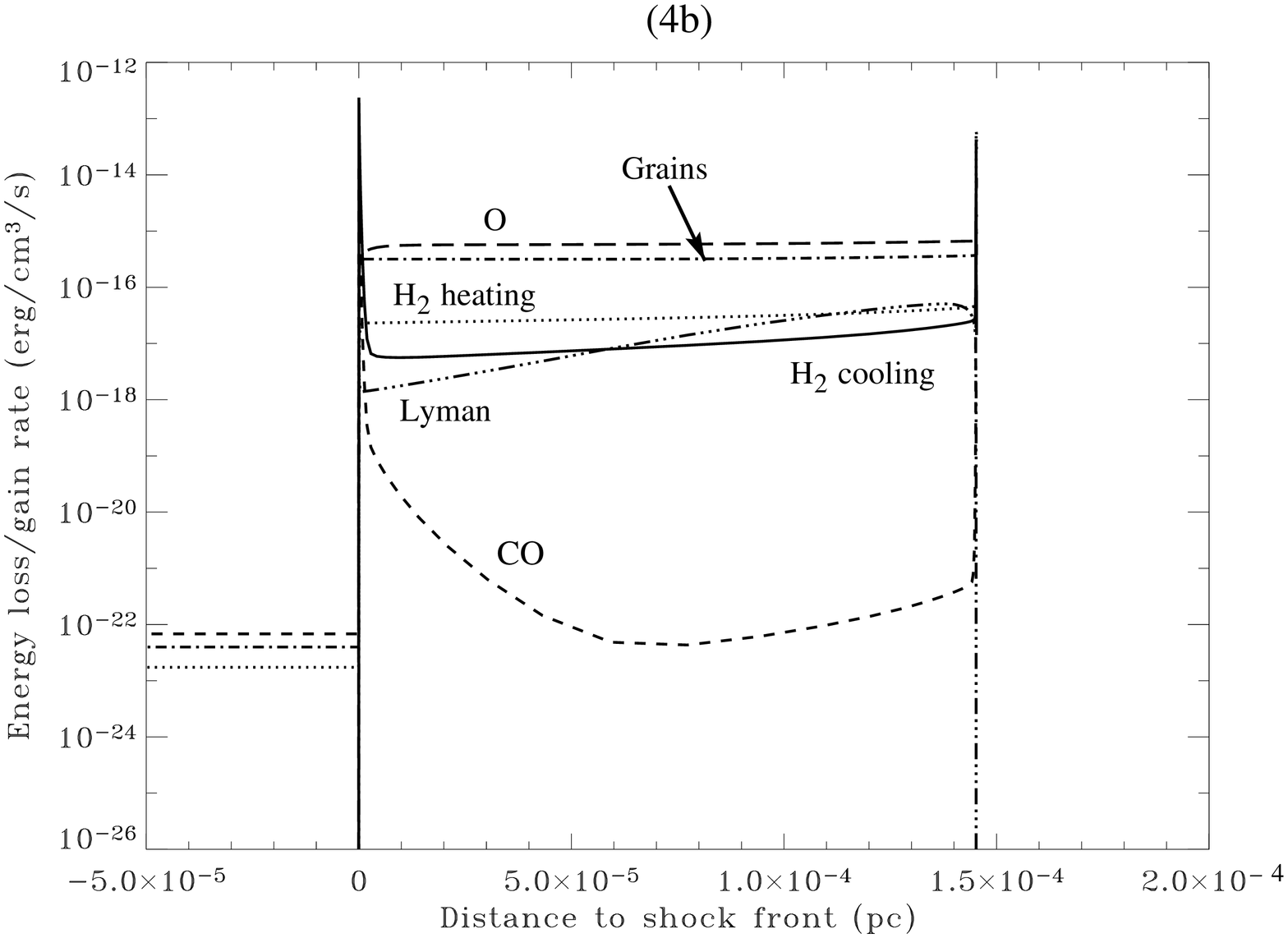,width=6cm,angle=-90}&
\psfig{file=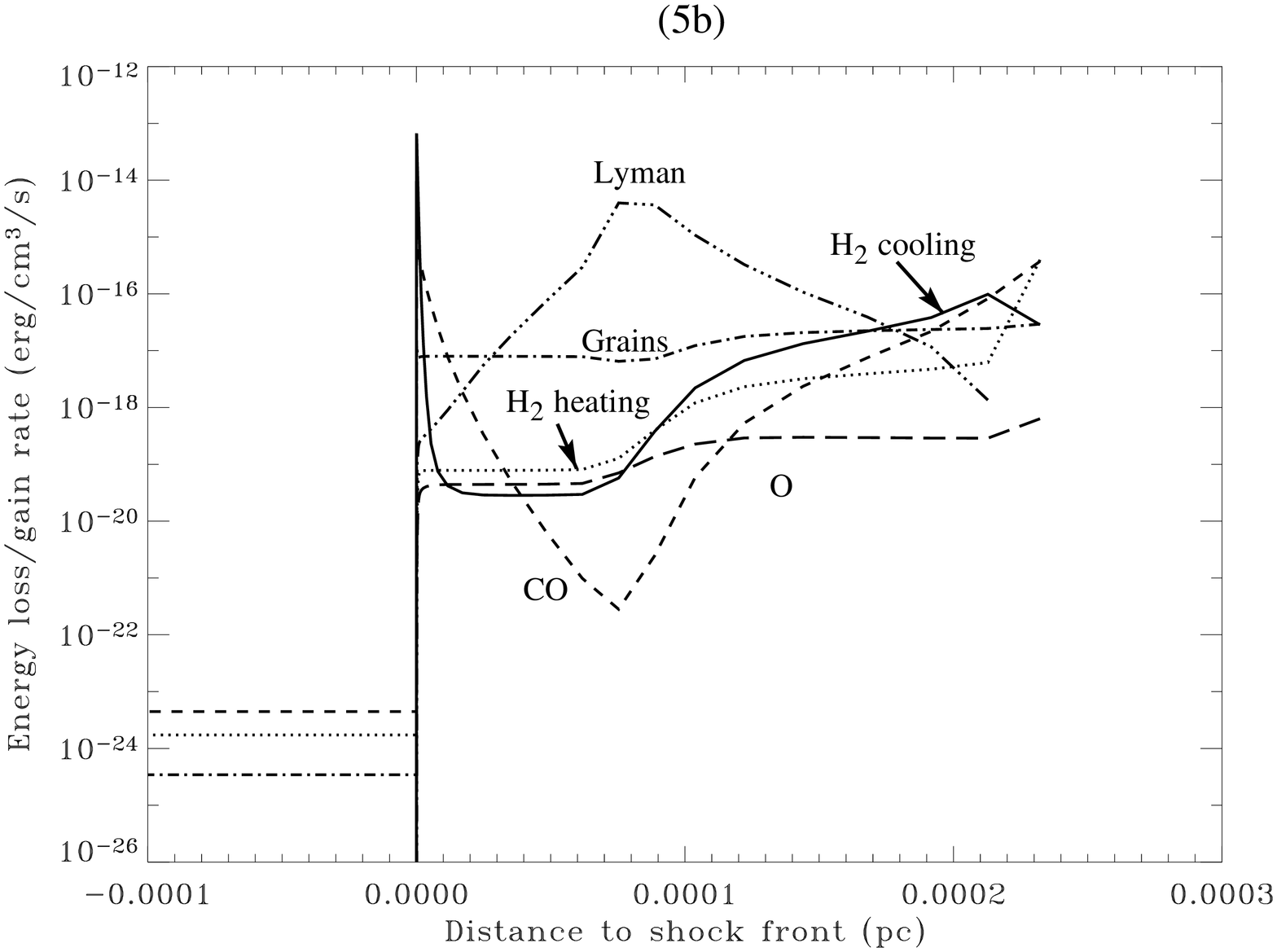,width=6cm,angle=-90}\\
\psfig{file=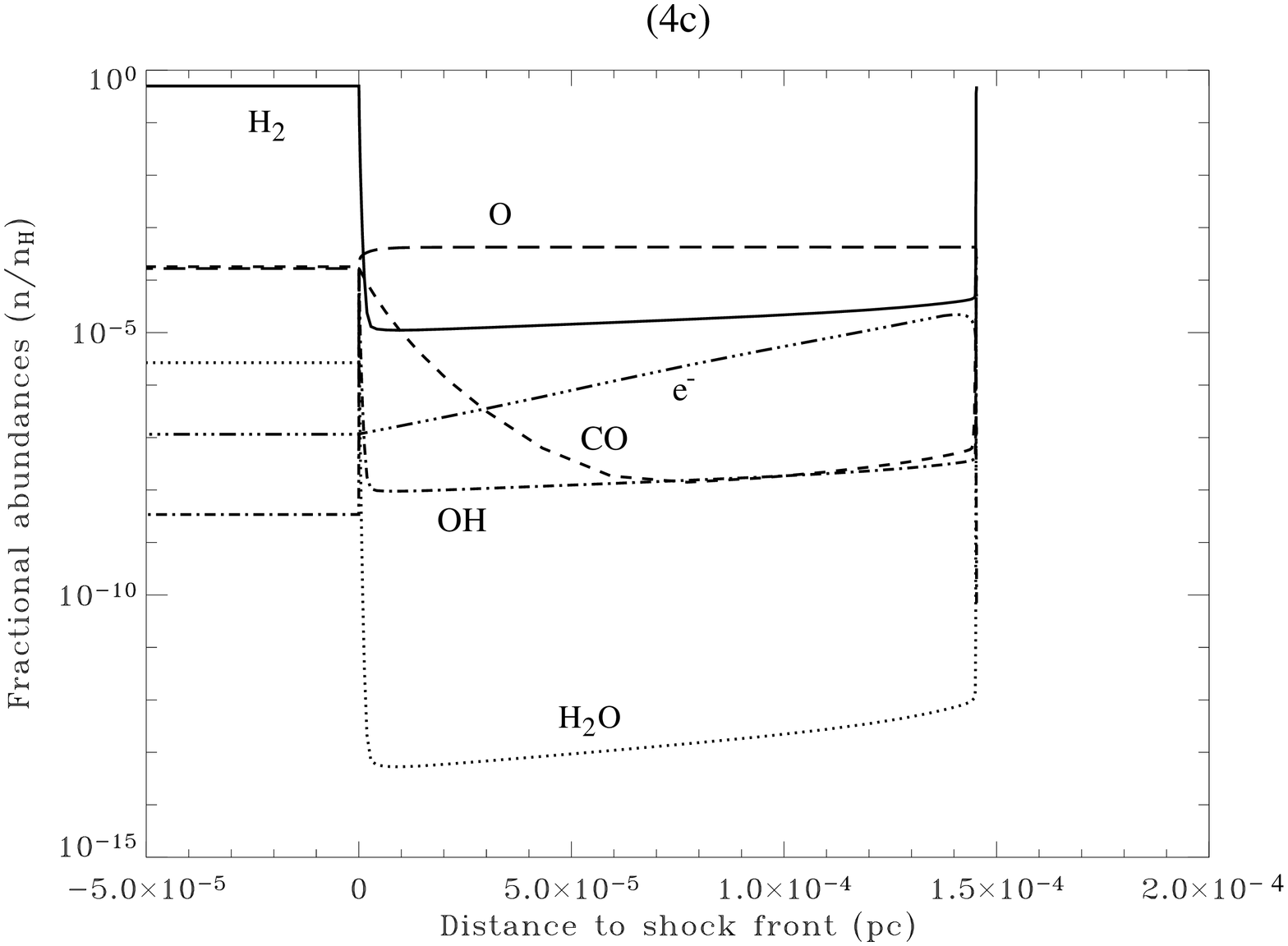,width=6cm,angle=-90}&
\psfig{file=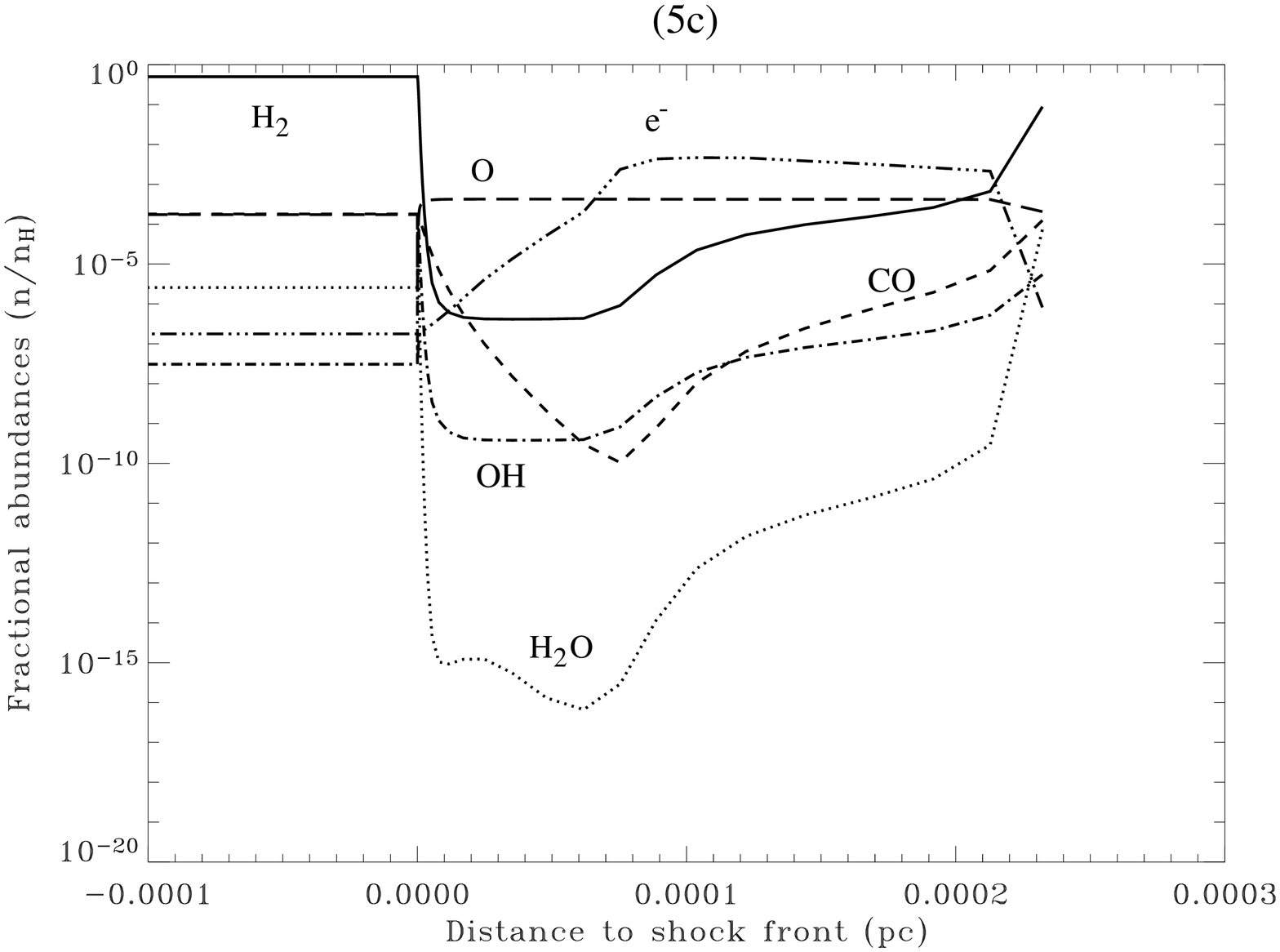,width=6cm,angle=-90}\\
\psfig{file=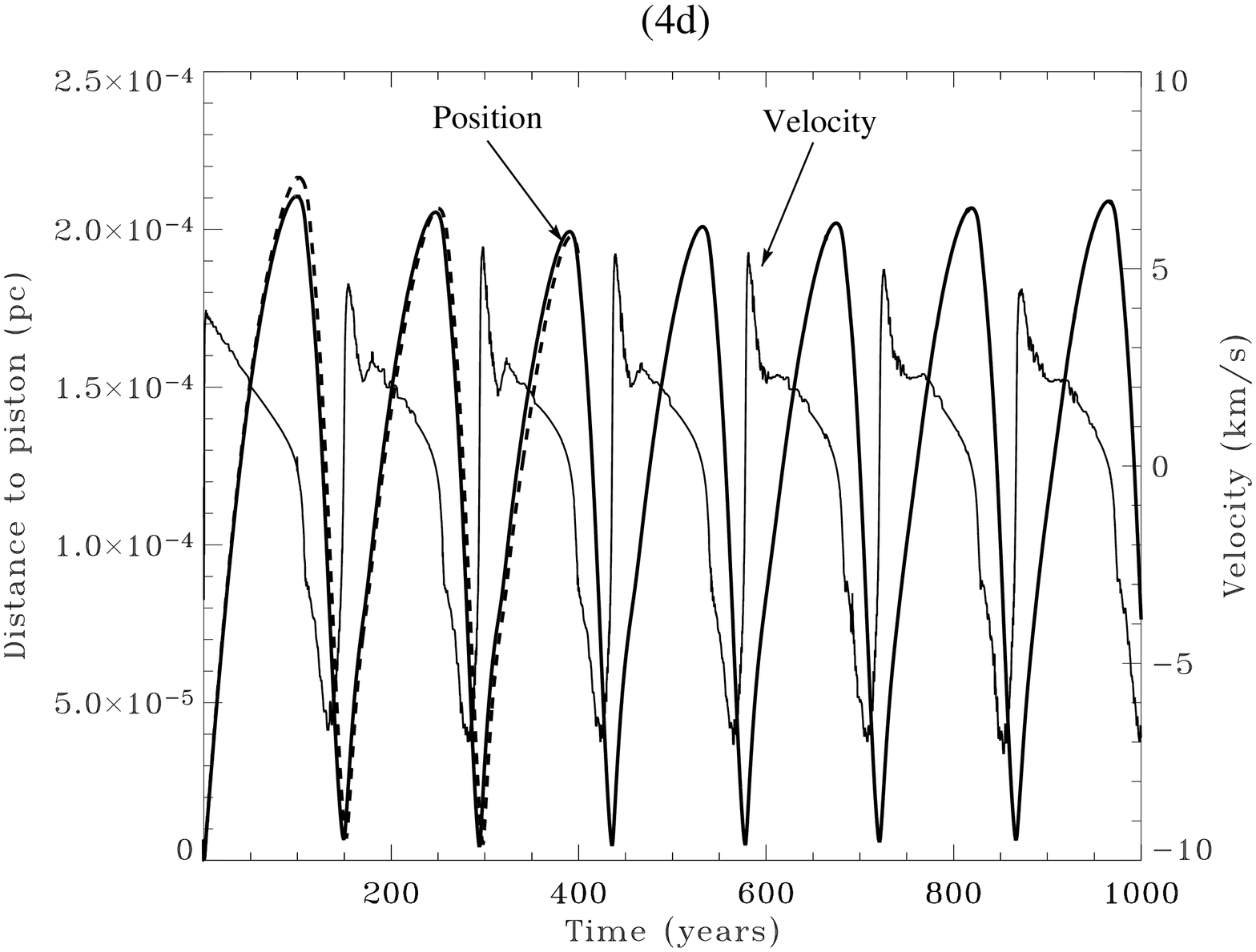,width=6cm,angle=-90}&
\psfig{file=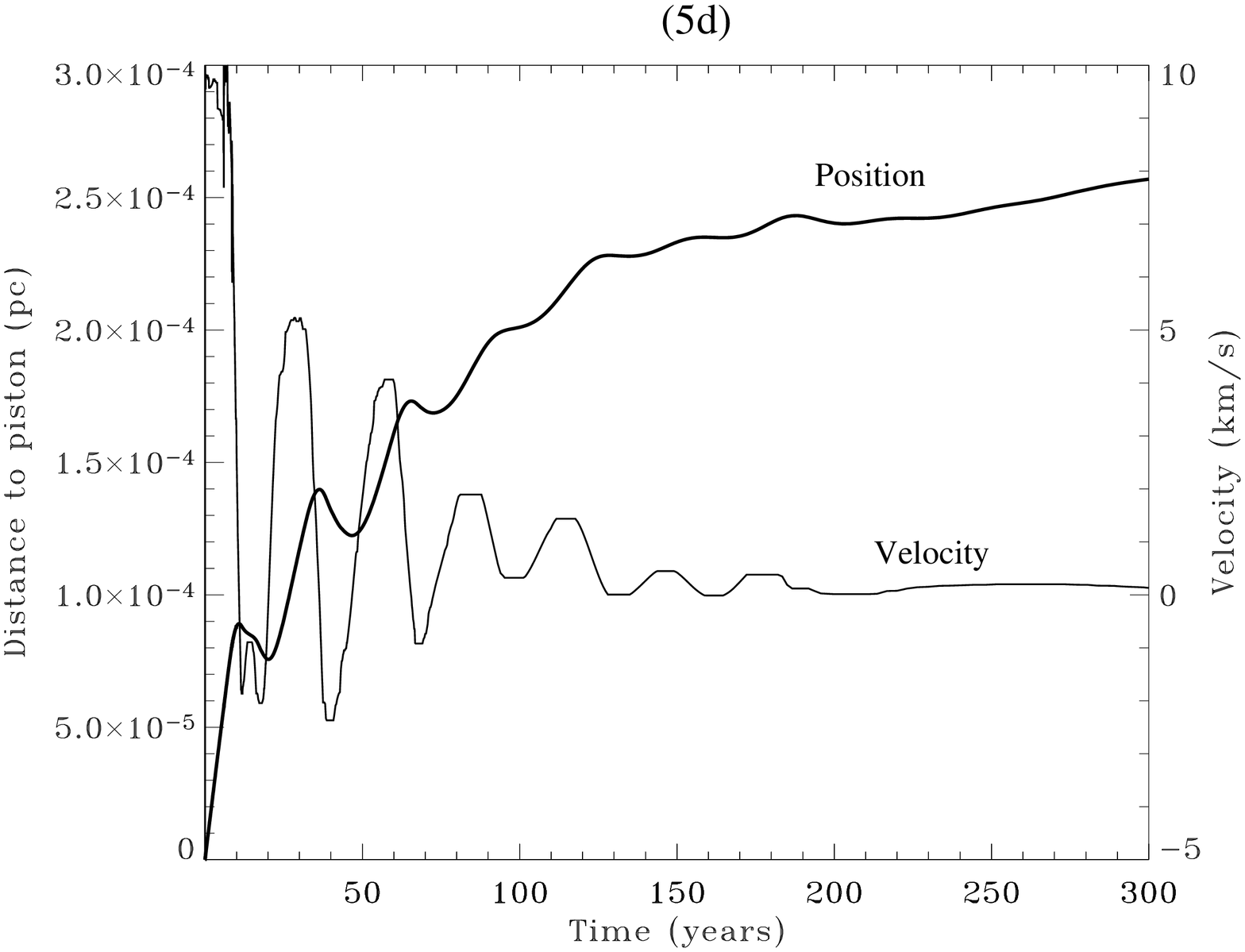,width=6cm,angle=-90}
\end{tabular}
}
\caption{Same as Fig.~1 for a weakly dissociative
J-shock of parameters $n=10^5~$cm$^{-3}$, $u=25$~kms$^{-1}$, $b=0$ at $t=50$~yr.
In panel (d), the thick solid line is the trajectory of the shock in a
simulation with a reduced network of 8 species. The simulation with the
whole network is plotted in a thick dashed line, up to the point where it
stalled. 
{\bf Fig.~5.} Same as Fig.~1 for a partly ionising
J-shock of parameters $n=10^4~$cm$^{-3}$, $u=40$~kms$^{-1}$, $b=0$ at $t=220$~yr. 
}
\label{Jplot}
\end{figure*}
\addtocounter{figure}{1}

At the bottom of these figures, we also plot the shock trajectories,
i.e. the positions as a function of time of the C precursor and the J
front (defined as the zones with maximum ratio of viscous to thermal
pressure in the ionic and neutral fluids, respectively). We also plot their
corresponding propagation velocity $v$ away from the piston, obtained by
differentiating with respect to time, and smoothing with a median filter
(20 time steps of width). This avoids spurious high velocities due to
change in the zone number.

 Even at intermediate times, $v$ for stable J-type fronts is found
to obey the steady-state relation $v=u/(C-1)$ where $C$ is the neutral
fluid compression factor at the piston. In the initial adiabatic
phase, $v=u/3$ ($C=4$ for $\gamma=5/3$). It then decreases during the
cooling phase, down to a steady value when thermal equilibrium is
reached in the post-shock gas. These three phases can be very clearly
seen in Fig.~1d. For the C-type precursor,
the initial velocity corresponds to the ion magnetosonic speed. $v$
then decreases as $u/(C-1)$ where $C$ is now the magnetic compression
factor \citep[see][ hereafter Paper~II]{Les04}. At early times, the J-type feature of a
magnetised shock behaves like the same shock without magnetic field,
as neutrals and ions are still decoupled. Thus, the J-front propagates
much slower than the C-front, which allows the magnetic precursor to
grow. In the steady C-type case, the J-front velocity eventually
catches up with the C-front due to the softening of its entrance
conditions. In the steady CJ-type case, the C-front velocity is forced
to slow down to the J-front value by early effective recoupling
between the neutrals and the magnetic field through their collisional
interactions with ions, and the precursor growth is halted.

\subsection{Time scales and length scales}
Table \ref{timescales} gives, next to the final state of the shock
(J, CJ, or C), the time scales in years taken to reach
steady state. It includes in parentheses the width (in pc) of
the steady-state shock structure (we give the length of both C and J
components for CJ-type shocks). 

  We compared a time evolution sequence for parameters $b=1$,
  $n=10^3$, and $u=10$~kms$^{-1}$ to Fig.~6 of \citet{CPF98}. We find a
  very good agreement,  though we use a very different method of
  integration, that cooling processes have been strongly refreshed
  to account for dissociative shocks, and that our resolution is higher at both
  shocks and relaxation layer. The  method
  used by \citet{CPF98} to recover the time is hence validated.
 
 In particular, we confirm the long time scales to steady C-type
 shocks found by these authors, of the order of the ion crossing time
 scale, and their weak dependence on the magnetic field for $0<b \le
 1$. We add here (see Table~\ref{timescales}) that these time scales
 are roughly inversely proportional to the density, as expected from
 the cooling time scales, and that there is no significant change with
 respect to the velocity, except when dissociative and
 non-dissociative velocities are compared \citep[see also Fig.~4
 of][ for example ]{L02}. The length scales of the shocks which
 actually dissociate molecules are longer than the non-dissociating
 ones by a factor of 5-10, due to the fact that H$_2$ is a very efficient
 coolant.

\label{Cshocks}

\begin{table*}[h]
\centering

\begin{tabular}{clrlrlr}
\hline
\hline

&&&&&&\\
$b=0$&\multicolumn{2}{c}{$n = 10^{3}~$cm$^{-3}$}&\multicolumn{2}{c}{$n = 10^{4}~$cm$^{-3}$}&\multicolumn{2}{c}{$n = 10^{5}~$cm$^{-3}$}\\
   $u$& $t$ (yr)& size (pc) &$t$ (yr)& size (pc) &$t$ (yr)& size (pc) \\
\hline
&&&&&&\\
$10$~kms$^{-1}$&J=10$^4$ & (10$^{-4}$)  &J=2~10$^3$ & (2.5~10$^{-5}$)&J=2~10$^2$ & (6~10$^{-6}$)\\
&&&&&&\\

$20$~kms$^{-1}$&J=10$^4$ & (5~10$^{-5}$)&J=2~10$^3$ & (10$^{-5}$)    &J=2~10$^2$ & (4~10$^{-6}$)\\
&&&&&&\\

$25$~kms$^{-1}$&         &           &J=5~10$^{2}$& (1.5~10$^{-5}$)& Bouncing &            \\
         &         &           &            &              & A=150 & (2.2~10$^{-4}$)\\
&&&&&&\\

$30$~kms$^{-1}$&J=10$^4$& (4~10$^{-5}$)&Bouncing &                  &Oscillating &           \\
         &        &            &A=700& (8~10$^{-4}$)       & &      \\
         &        &            &o=100& (5~10$^{-5}$)       & o=40 &(5~10$^{-5}$)                      \\
&&&&&&\\

$40$~kms$^{-1}$&Oscillating&          &J=2~10$^{2}$& (2.5~10$^{-4}$)&J=40& (2~10$^{-5}$)    \\
         &o=650& (4~10$^{-4}$)    &o'=40&(3~10$^{-5}$)         &o'=2.5&(2~10$^{-6}$)    \\
&&&&&&\\
\end{tabular}

\medskip
\begin{tabular}{clrlrlr}
\hline
\hline

&&&&&&\\
$b=0.1$&\multicolumn{2}{c}{$n = 10^{3}~$cm$^{-3}$}&\multicolumn{2}{c}{$n = 10^{4}~$cm$^{-3}$}&\multicolumn{2}{c}{$n = 10^{5}~$cm$^{-3}$}\\
   $u$& $t$ (yr)& size (pc) &$t$ (yr)& size (pc) &$t$ (yr)& size (pc) \\
\hline
&&&&&&\\
$10$~kms$^{-1}$&C=10$^5$&(2~10~$^{-2}$)     &C=10$^4$&(3~10$^{-3}$) &C=1.5~10$^3$&(4~10$^{-4}$) \\
&&&&&&\\

$20$~kms$^{-1}$&C=10$^5$&(2~10~$^{-2}$)     &C=10$^4$&(3~10$^{-3}$) &C=1.5~10$^3$&(4~10$^{-4}$) \\
&&&&&&\\

$30$~kms$^{-1}$&C=10$^5$ & (2~10~$^{-2}$)   &C$>$700           &        &CJ=10$^{2}$&(C=2~10$^{-5}$)\\
         &         &         &A'=700  &(8~10$^{-4}$)   &       &(J=3~10$^{-4}$)\\
         &         &        &o'=150   &(5~10$^{-5}$)   &o=20&(10$^{-5}$)           \\ 
&&&&&&\\

$40$~kms$^{-1}$&CJ=5~10$^3$ &(C=6~10$^{-5}$)  &CJ=250   &(C=10$^{-5}$)  &CJ=40 &(C=10$^{-4}$)  \\
         & & (J=$5~10^{-3}$)   & & (J=10$^{-3}$)         && (J=10$^{-4}$)             \\
         &o=500& (3~10$^{-4}$)&o=25& (3~10$^{-5}$)     &o=3 &(3~10$^{-6}$)          \\ 
&&&&&&\\
\end{tabular}

\medskip
\begin{tabular}{clrlrlr}
\hline 
\hline

&&&&&&\\
$b=1$&\multicolumn{2}{c}{$n = 10^{3}~$cm$^{-3}$}&\multicolumn{2}{c}{$n = 10^{4}~$cm$^{-3}$}&\multicolumn{2}{c}{$n = 10^{5}~$cm$^{-3}$}\\
   $u$& $t$ (yr)& size (pc) &$t$ (yr)& size (pc) &$t$ (yr)& size (pc) \\
\hline
&&&&&&\\
$10$~kms$^{-1}$&C=10$^5$&(2~10~$^{-1}$)     &C=1.3~10$^4$&(4~10~$^{-2}$) &C=1.5~10$^3$&(6~10$^{-3}$)\\
&&&&&&\\

$20$~kms$^{-1}$&C=10$^5$&(2~10~$^{-1}$)     &C=1.3~10$^4$&(4~10~$^{-2}$)  &C=1.5~10$^3$&(5~10$^{-3}$)\\
&&&&&&\\

$30$~kms$^{-1}$&C=10$^5$&(2~10~$^{-1}$)     &C=10$^4$&(3~10~$^{-2}$)      &C=1.5~10$^3$&(4~10$^{-3}$)\\
         &   &               &A'=1000&(10$^{-3}$)  &A'=500&(4~10$^{-4}$)      \\
         &   &              &o'=150&(5~10$^{-5}$)&o'=15&(1~10$^{-6}$)       \\
&&&&&&\\

$40$~kms$^{-1}$&C=4~10$^4$&(10~$^{-1}$) &CJ=10$^3$&(C=6~10$^{-3}$)  &CJ=2~10$^2$&(C=10$^{-3}$)\\
         &A'=4000&(3~10$^{-3}$)&&(J=5~10$^{-2}$)               &&(J=10$^{-2}$)           \\
         &o'=600&(10$^{-4}$) &o'=25&(3~10$^{-5}$) &o'=2.5&(10$^{-6}$)      \\
&&&&&&\\
\end{tabular}
   
\caption{For each run, we provide the final steady-state of each
shock (J, CJ, C) followed by the time scale in years at which
steadiness is achieved.
The total width of the shock at this point (in pc) is given in
parentheses (in the case of CJ-type shocks, J is the length of the
relaxation layer, and C is the length of the magnetic
precursor). C$>700$ indicates that a J-front is still present when the
code stalls at $t=700$~yr.  A and o are the average duration of one
large arch or one small oscillation, when present. The associated
length scales correspond to the typical amplitude of these
oscillations. A' stands for a lone arch. o' stands for damped small
oscillations.  }
\label{timescales}

\end{table*}

\subsection{Conditions for steady CJ-type shocks}
 
When the magnetic field is lower than a critical value, the {\it
steady} shock is composed of a magnetic precursor, followed by a
J-shock behind which the two fluids recouple, and the gas relaxes
towards its final post-shock state. \citet{CPF98} have already shown
one such model for non-dissociative velocities in a diffuse medium
(for $n_{\rm H}=25$~cm$^{-3}$, $u=10$~kms$^{-1}$ and $B=5~\mu$G). This
work presents models for dissociative and partly ionising velocities,
which were out of reach of their anamorphosis method, because it could
not resolve {\it both} the J-front and the H$_2$ reformation zone. The
final steady structure of such a shock can be seen in Fig.~2. It mixes
all the characteristics of a C-type precursor with those of a partly
ionising J-type shock (see Fig.~5).

  \citet{L02} computed the critical values of the magnetic field and
  velocity for the transition from C to CJ-type behaviour. Their
  criterion to determine when a J front would remain in a stationary
  shock with a magnetic field is the presence of a sonic point in the
  neutral velocity profile (in the shock frame). Such a sonic point is
  not encountered in any of our CJ-type models. Hence, the criterion
  previously used for CJ transition is too strong, and the range of
  parameters where stationary C-type shocks exist should be reduced
  compared to their results.

In Paper~II, we give a means of determining the fate (C or CJ) of a
magnetic shock with the only help of a steady-state code for
non-dissociative velocities.

\subsection{Chemically driven oscillations}
\label{Jdiss}

We have identified two (non mutually exclusive) situations where
oscillations occur~: weakly dissociative shocks which undergo large
rebounds driven by H$_2$ reformation, and partly ionising shocks which show
smaller oscillations probably driven by H ionisation.

\subsubsection{Oscillations due to H$_2$ dissociation/reformation} 

\begin{figure}[h]
\begin{center}
\begin{tabular}{c}
\psfig{file=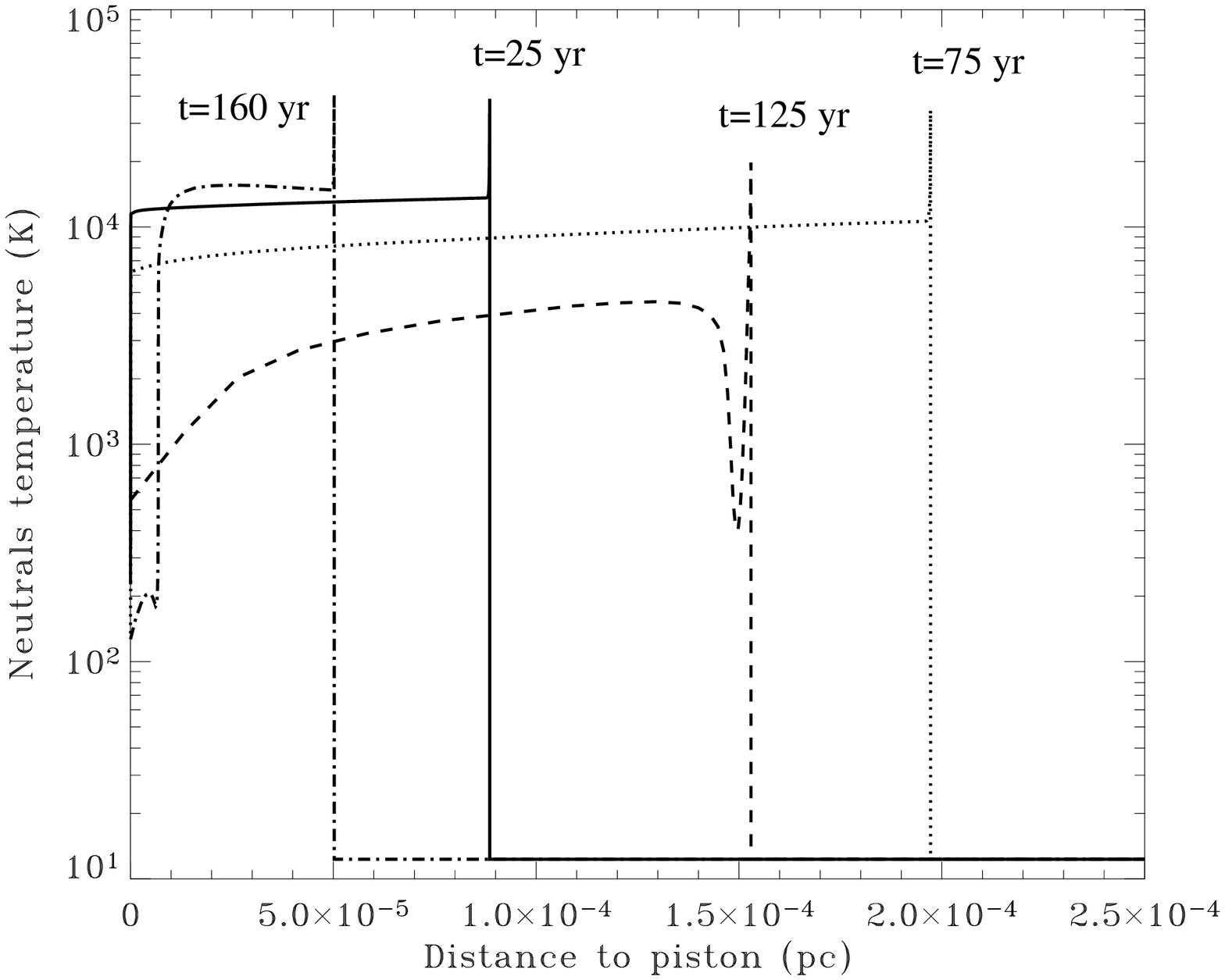,angle=-90,width=8.8cm}\\
\psfig{file=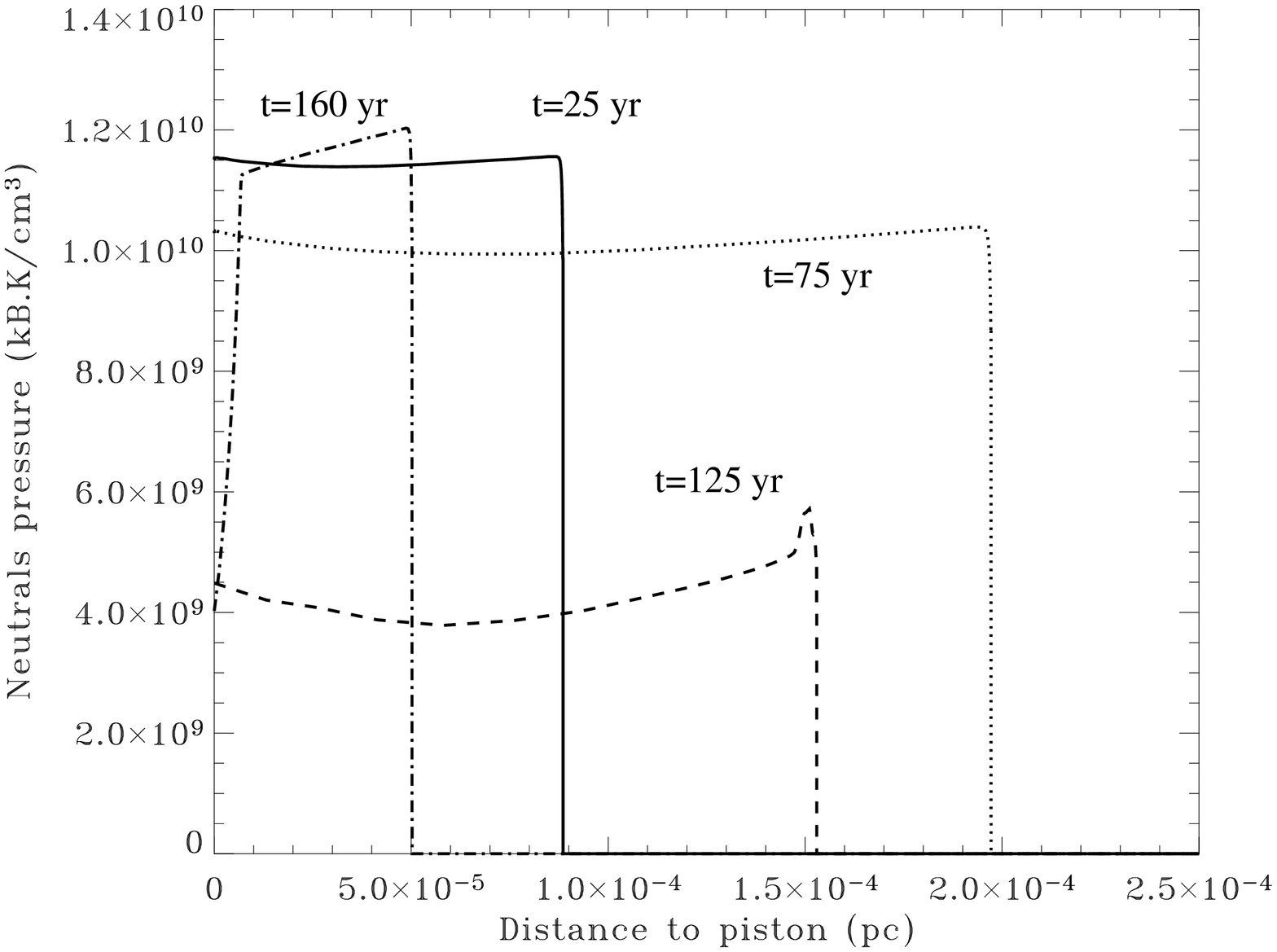,angle=-90,width=8.8cm}
\end{tabular}
\end{center}
\caption{Temperature and pressure snapshots for the first period
of the bouncing shock with parameters $b=0$, $n=10^5~$cm$^{-3}$, and
$u=25$~kms$^{-1}$.  The plotting scale for the temperature is logarithmic
while it is linear for the pressure.  }
\label{period}
\end{figure}

We find a narrow range of velocities, close to the dissociation limit $u_{\rm d}$,
where shocks cycle between an expansion phase where the shock is
dissociative, and a retreat phase where the shock is non-dissociative (an
example is shown in Fig.~4).
Even if $u<u_{\rm d}$, it can happen that the initial entrance velocity
in the shock $u^0=u+v$ is greater than $u_{\rm d}$\footnote{Because the compression
factor in the atomic plateau after a fully dissociative shock is around 6
(see Paper~II), this happens for $u>\frac56u_{\rm d}$}.  We define such shocks as
weakly dissociative shocks.  Fig.~\ref{period} shows the first period of
the shock with parameters $b=0$, $n=10^5~$cm$^{-3}$ and $u=25~$kms$^{-1}$.

In the dissociative expansion phase, the front is followed by an
isothermal atomic plateau at $T\simeq$10$^4$~K. The length of the plateau
is usually governed by O cooling followed by H$_2$ reformation. When H$_2$
reaches a high enough abundance, it becomes again the main coolant. At this
point, the higher the H$_2$ abundance, the faster it cools and is
compressed, and the faster it is reformed. This run-away process yields a
very sharp exit from the plateau (see Fig.~4a).

When the compression factor at the end of the plateau increases, $u^0$
gets closer to $u$. Therefore, $u^0<u_{\rm d}$, and H$_2$ survives the
shock. Fast cooling ensues, and the shock collapses until it rebounds
near the piston, with a total thickness on the order of the H$_2$
cooling length. The cycle is repeated and the shock undergoes large
oscillations (see Fig.~4d). Here, the H$_2$
reformation instability proceeds exactly like the thermal instability
of the 150~kms$^{-1}$ model of \citet{G88} or the type C (as defined by
them) models of \citet{WF96}~: the gas condenses isobarically in a
shell as shown on Fig.~\ref{period}.

To assess whether the bounces (``arches'') are persistent or damped in
the $b=0$ case, we computed at least ten periods with the help of a reduced
network of 8 species (see Fig.~4d). The characteristic periods
(in years) and typical amplitudes (in pc) of the arches (A) are given
in table
\ref{timescales}. These time and length scales are found to depend strongly
on the parameters $n$ and $u$.  They also depend on the control
parameters of the grid, as well as on properties of the cooling functions
(like the collisional excitation of CO by H), although the qualitative features
remain unchanged. 


For weakly dissociative shocks with non-zero $b$, the J-type feature
initially behaves like the same shock without magnetic field, but the
magnetic precursor rapidly decelerates the neutrals velocity below the
threshold for dissociation, and further rebounds are switched off. Hence we
observe only one arch when a magnetic field is present. The period (in
years) and amplitude (in pc) of this first arch (A') are given in table
\ref{timescales}. They are very similar to their values for $b=0$.

\subsubsection{Oscillations due to H ionisation}

For higher velocities, temperatures are sufficient to start to
collisionally ionise H in the plateau.  As a result, the electron
fraction increases progressively. If the ionisation fraction is high
enough (i.e. if $u^0$ is large enough) Lyman cooling becomes dominant
over Oxygen, and the end of the plateau is first triggered by this
cooling. Once the temperature gets below 10$^4$ K, Lyman cooling drops
and the gas enters a second plateau. The end of the second plateau is
triggered by run-away H$_2$ reformation as previously described.  An
example of the thermal, chemical, and cooling structure of such a
shock is shown in Fig.~5. We define shocks that present two such
plateaux as "partly ionising shocks".

In these shocks, the length of the first plateau is governed by the
ionisation length of H. Since the temperature $T_{\rm p}$ inside the
first plateau is quadratic in $u$ (see equation 29 of Paper~II), and
the ionisation length is exponential in $1/T_{\rm p}$, there is a very
strong dependence of this length on the entrance velocity. This strong
dependence seems to drive an oscillatory behaviour (denoted "o" in
Table~1), with an amplitude on the order of the ionisation length. It
appears to be damped over a few periods in most cases.  Figure~5d
illustrates this temporal behaviour.

Partly ionising shocks are not exclusive of weakly dissociative
shocks, and a few shocks present both characteristics (those marked
with both an "A" and an "o" in table \ref{timescales}).

The relaxation layer behind a partly ionising J-front with $b \neq 0$ is
very similar to pure J-type shocks (compare Fig.~2 and Fig.~5).
 The only difference is the onset of a typical
density where thermal pressure is relayed by magnetic pressure. At this
point, and for sufficiently late times, the compression stops. This makes
the length scales of the H$_2$ reformation layer behind this point much
greater, but the thermal and chemical evolutions are quite the same, as
well as the oscillation properties.

\subsubsection{Previously known oscillations of shocks with chemistry}

\citet{Lim02} and \citet{SR03} also encounter oscillations in their
simulations of time-dependent shocks with chemistry. We compare our
findings to their work.

 \citet{Lim02} use explicit adaptive mesh refinement with chemistry
 splitted from hydrodynamics. They also follow H ionisation and H$_2$
 dissociation in a time-dependent way. However, they do not report any
 instability in the $n=10^4$ and $u=30~$kms$^{-1}$ case, while we see strong
 bouncing oscillations. This difference could arise because they do
 not model H$_2$ reformation on the grains, but only the reformation
 via the H$^-$ process, which is much slower. They do identify
 various oscillating behaviours in the course of their accelerated
 low-density shock, including oscillations of small
 amplitude and short time scales related to the ionisation of H when
 $u \simeq 50$ km~s$^{-1}$, which may be similar to ours. However,
 their results are difficult to compare to our constant inflow
 speed, much higher density simulations.

\citet{SR03} present shock simulations with the same boundary conditions
and densities as ours. Their inflow velocities are too high ($u \ge 40$ km
s~$^{-1}$) to observe the dissociation/reformation bouncing
oscillation, since their shocks are always dissociative.  However, they
observe a ``quasi-periodic or chaotic collapse'' driven by CO reformation
for a very wide range of velocities ($u$ from 40 to 60~kms$^{-1}$). The main
differences with our work are their approximation of chemical equilibrium
and optically thin emission for CO which makes CO cooling dominant in the
atomic plateau, along with their coarser time-step control and 
lower resolution.

\section{Discussion}

 We emphasise here a few points that remain to be investigated.

The opacity of the CO, H$_2$O, and OH molecular coolings could be more
detailed, including an LVG treatment in the rapidly compressing parts of
the shocks. This would increase their importance compared to H$_2$, though
perhaps not enough to trigger the large instabilities found in the
dissociative shock simulations of \citet{SR03}.  Even at our extremely
high resolution, we have also observed an effect of the grid parameters on
the amplitude and period of oscillations. A linear analysis remains to be
carried out to determine accurate periods without the caveats of the
numerics.

Next, the reformation of molecules in the relaxation layer showed the
necessity of a treatment of the chemical diffusion processes. We did not
check the influence of the diffusion model on the reformation zone.
Taking the diffusion of heat into account might change the thermal
behaviour of our shocks.  At higher velocities, the Lyman cooling and
probably a few more atomic coolings will not be optically thin anymore and
should also be treated more carefully.

  As \citet{L02} pointed out with steady models, the time-dependent
   tracking of the populations of H$_2$ has a large impact on the
  dissociation and hence the dynamical behaviour of shocks.  This
  effect deserves more investigation in a fully time-dependent context.

 Oblique magnetic fields and grain dynamics have
been completely neglected in this study. The lack of a velocity for the
grains may prove to be the main caveat of our study as preliminary
time-dependent simulations and steady-state calculations suggest
\citep{CR02,FP03}. In a steady-sate context, \citet{PH94} proved the
dynamical importance of oblique magnetic fields, although it has not
yet been investigated in any multifluid, time-dependent
studies. However, our
numerical technique should prove useful to model both these
effects.

\section{Conclusions}

  The adaptive moving grid  technique proves to be a unique
  tool to model time-dependent magnetised molecular shocks. It is
  currently the only algorithm able to model time-dependent
  dissociative shocks in presence of a magnetic field. With this tool,
  we validate previous results by \citet{CPF98} and we investigate the
  formation of shocks in the conditions of protostellar jets.

  We find in agreement with \citet{P97,SM97,CPF98} that C-shocks are steady
  after a very long time delay. We produce the first C-type and
  CJ-type shocks in the dissociative range of velocities.  We point out
  that the occurrence of a sonic point is not a valid criterion for the
  transition to CJ-type steady states.

In our simulations, we also discover two oscillatory behaviours, which
are linked to H$_2$ dissociation/reformation and to Hydrogen
ionisation, respectively. The oscillations vanish when a strong
magnetic field is present and after a significant magnetic precursor
has built up.  The oscillation periods and amplitudes are found to
depend strongly on the density and inflow speed. A detailed treatment
of CO opacity is required to assess the reality of the CO-driven
instabilities in dissociative shocks reported by \citet{SR03}.

  In a companion paper (Paper~II), we analyse
  our results in comparison with truncated steady state models. We
  derive analytical relations as well as constructions for
  intermediate times of non-dissociative shocks with or without
  magnetic fields.

\begin{acknowledgements}
We thank the referee (Pr. T.W. Hartquist) for his 
critical remarks on this paper, which lead us to clarify its scope and
contribution to the field, and to greatly improve its presentation. 
\end{acknowledgements}

\bibliographystyle{aa}

\end{document}